\documentclass[longauth]{aa}
\usepackage[varg]{txfonts}
\pdfoutput=1 
\usepackage{breqn}
\usepackage{natbib}
\defcitealias{SCpaper}{W14}
\defcitealias{Courteau1997}{C97}
\defcitealias{Pizagno2007}{P07}
\bibpunct{(}{)}{;}{a}{}{,} 
\usepackage{flexisym}
\usepackage{textcomp}
\usepackage{enumerate}
\usepackage[bookmarks=false]{hyperref}
\usepackage{listings}
\usepackage{url}

\begin{document}

\title{The space density distribution of galaxies in the absolute magnitude - rotation velocity plane: a volume-complete Tully-Fisher relation from CALIFA stellar kinematics}

\institute{
Leibniz-Institut f\"ur Astrophysik Potsdam (AIP), An der Sternwarte 16, D-14482 Potsdam, Germany\label{i:AIP} 
\and
Dept. Astrof\'{i}sica, Universidad de La Laguna, C/ Astrof\'{i}sico Francisco S\'anchez, E-38205-La Laguna, Tenerife, Spain\label{i:IAC}
\and
Instituto de Astrof\'{i}sica de Canarias, C/V\'{\i}a L\'actea S/N, 38200-La Laguna, Tenerife, Spain\label{i:ULL}
\and
Max Planck Institute for Astronomy, K\"onigstuhl 17, D-69117 Heidelberg, Germany \label{i:MPIA}
\and
Instituto de Astronom\'\i a, Universidad Nacional Auton\'oma de Mexico, A.P. 70-264, 04510, M\'exico,D.F.  \label{i:unam}
\and 
Instituto de Astrof\'{i}sica de Andaluc\'{i}a (CSIC), Glorieta de la Astronom\'{\i}a, s/n, 18008 Granada, Spain \label{i:IAA}
\and
Centro Astron\'{o}mico Hispano Alem\'{a}n, Calar Alto, (CSIC-MPG), C/Jes\'{u}s Durb\'{a}n Rem\'{o}n 2-2, E-04004 Almer\'{\i}a, Spain\label{i:CAHA}
\and
Department of Physics, Royal Military College of Canada, P.O. Box 17000, Station Forces, Kingston, ON, K7K 7B4, Canada\label{i:RMCC}
\and
Kapteyn Astronomical Institute, Rijksuniversiteit Groningen, Postbus 800, NL-9700 AV Groningen, The Netherlands \label{i:Kapteyn} et al.
\and
University of Vienna, T\"urkenschanzstr. 17, 1180 Vienna, Austria \label{i:vienna}
\and
Institute of Astronomy, School of Physics, University of Sydney, NSW 2006, Australia \label{i:SydUni}
\and
Departamento de Astrof\'{i}sica y CC. de la Atm\'{o}sfera, Universidad
Complutense de Madrid, E-28040, Madrid, Spain \label{i:UCM}
}

\author{S.~Bekerait{\.e}, \inst{\ref{i:AIP}}; C.J.~Walcher\inst{\ref{i:AIP}}; J.~Falc{\'o}n-Barroso\inst{\ref{i:IAC}, \ref{i:ULL}}; B.~Garcia Lorenzo\inst{\ref{i:IAC}}; M.~Lyubenova\inst{\ref{i:MPIA},\ref{i:Kapteyn}}, S.F.~S\'anchez\inst{\ref{i:IAA},\ref{i:CAHA},\ref{i:unam}}; K.~Spekkens\inst{\ref{i:RMCC}}; G.~van de Ven\inst{\ref{i:MPIA}}; L.~Wisotzki\inst{\ref{i:AIP}}; B.~Ziegler\inst{\ref{i:vienna}}, J.A.L.~Aguerri\inst{\ref{i:IAC},\ref{i:ULL}}; J.~Barrera-Ballesteros\inst{\ref{i:IAC},\ref{i:ULL}}, J.~Bland-Hawthorn\inst{\ref{i:SydUni}}, Cristina Catal\'{a}n-Torrecilla\inst{\ref{i:UCM}}; R.~Garc\'{i}a-Benito\inst{\ref{i:IAA}}; the CALIFA collaboration}

\date{Received date / Accepted date }

\abstract{We measure the distribution in absolute magnitude - circular velocity space for a well-defined sample of 199 rotating Calar Alto Legacy Integral Field Area Survey (CALIFA) galaxies using their stellar kinematics. 
Our aim in this analysis is to avoid subjective selection criteria and to take volume and large-scale structure factors into account. Using stellar velocity fields instead of gas emission line kinematics allows including rapidly rotating early type galaxies. Our initial sample contains 277 galaxies with available stellar velocity fields and growth curve $r$-band photometry. After rejecting 51 velocity fields that could not be modelled due to the low number of bins, foreground contamination or significant interaction we perform Markov Chain Monte Carlo (MCMC) modelling of the velocity fields, obtaining the rotation curve and kinematic parameters and their realistic uncertainties. We perform an extinction correction and calculate the circular velocity $v_{\mathrm{circ}}$ accounting for pressure support a given galaxy has. The resulting galaxy distribution on the $M_r$ - $v_{\mathrm{circ}}$ plane is then modelled as a mixture of two distinct populations, allowing robust and reproducible rejection of outliers, a significant fraction of which are slow rotators. The selection effects are understood well enough that the incompleteness of the sample can be corrected for and the 199 galaxies can be weighted by volume and large-scale structure factors enabling us to fit a volume-corrected Tully-Fisher relation (TFR). More importantly, we also provide the volume-corrected distribution of galaxies in the $M_r$ - $v_{\mathrm{circ}}$ plane, which can be compared with cosmological simulations. The joint distribution of the luminosity and circular velocity space densities, representative over the range of -20 $> M_r > $ -22 mag, can place more stringent constraints on the galaxy formation and evolution scenarios than linear TFR fit parameters or the luminosity function alone.
}

\authorrunning{Bekerait\.{e} et al. }
\titlerunning{The CALIFA Tully-Fisher relation}

\maketitle

\section{Introduction}

\subsection{The Tully-Fisher relation} 

The Tully-Fisher relation \cite[TFR,][]{Tully1977} links two intrinsic properties of rotationally supported galaxies: their circular rotation velocities and luminosities. Stated in physical terms, this relation indicates a close relationship between the total dynamical mass and the stellar mass \citep[or the total baryonic content, ][]{McGaugh2000} of the galaxies.

Circular velocities and luminosities of galaxies have long been used to estimate extragalactic distances \cite[see][for the first use]{Oepik1922}, also see \citet{Roberts1969, Bottinelli1971, Balkowski1974, Shostak1975} for early analyses of scaling relations of spiral galaxies.

Low intrinsic scatter of the TFR cannot be explained by initial conditions \citep{Eisenstein1996} which implies that the subsequent evolutionary processes were crucial in determining the shape of the relation. The fact that the TFR exists is thought to be a natural outcome of hierarchical structure assembly \citep{steinmetz1999}.

Although the TFR has primarily been envisaged and successfully used as a tool for extragalactic distances determination \citep{Tully1977}, it also offers fundamental insights into the processes of disk assembly and evolution. We try to summarize the many uses of TFR in the following paragraphs.

The TFR in its initial form and its many variants (relations between different measures of rotational velocity and the stellar mass, total baryonic mass, absolute magnitude in different passbands) has been extensively employed as a constraint on galaxy formation and evolution models \citep{Koda2000, Croton2006, Dutton2009, Dutton2011, Tonini2011, McCarthy2012, Vogelsberger2013}. It also provided independent constraints on cosmological parameters \citep{Eisenstein1996, vandenBosch2000, Masters2006}, has been used to test the predictions of $\Lambda$CDM \citep{Blanton2008} and to characterise properties of dark matter haloes such as their concentration \citep{Dutton2011} and response to galaxy formation \citep{Dutton2009, Chan2015}. The TFR has also been used to put constraints on virial properties of barred/unbarred galaxies \citep{Courteau2003}, disk submaximality \citep{Courteau1999, Courteau2015}, to investigate the origin of S0 galaxies \citep{Neistein1999, Williams2009, Tonini2011}, test the universality of the initial mass function \citep{Bell2001, Dutton2011} and to infer the galaxy velocity function \citep{Gonzalez2000}.

In addition, TFR measurements at higher redshifts provided insights into mode of gas accretion at $z \approx 2.2$ \citep{Cresci2009}, stellar-to-dynamical mass ratio at $z \approx 3$ \citep{Gnerucci2011}, disk assembly timescales \citep{Miller2012}, evolution of bulgeless galaxies \citep{Miller2013}, "downsizing" effect \citep{Boehm2007} and luminosity evolution of rotating disks \citep{Ziegler2002, Puech2008, Miller2011}.

Reproducing the observed redshift evolution, slope, offset and intrinsic scatter of the TFRs is a standard test of cosmological simulations. It has been a long-standing problem of cosmological simulations \citep{steinmetz1999, vandenBosch2000, Koda2000, Cole2000, Eke2001, Croton2006, courteau2007, Dutton2011}, however, significantly remedied by a combination of more sophisticated feedback implementations, prescriptions for dark halo response and increased accuracy of cosmological parameters. 

Several studies that describe the convergence on reproducing the observed TFR are those of \citet{Dutton2007, Governato2007, Trujillo-Gomez2011, Tonini2011, McCarthy2012, Vogelsberger2014}. 
Semi-analytical models by \citet{Tonini2011} reproduce the TFR at higher redshifts but yield too bright values at z=0, probably due to uncertainties in star formation histories. Hydrodynamical zoom-in resimulations by \citet{McCarthy2012} reproduce the TFR for galaxies with $\log(M_{*}) < 10.7$, claiming that the turn-off at the higher mass end is due to the lack of AGN feedback prescription. \citet{Governato2007} employ N-body SPH simulations with supernova feedback to produce disk galaxies that lie on both $I$-band and baryonic TFRs. \citet{Trujillo-Gomez2011} use Bolshoi dark matter-only simulations and abundance matching to demonstrate that the luminosity-velocity relation and the baryonic TFR match the observed ones. The Illustris project implements a sophisticated feedback model that includes both stellar and AGN feedback, reporting slightly too high circular velocities in the $M_{*} - v_{\mathrm{circ}}$ relation \citep{Vogelsberger2014}. The goal of this study is to provide a measurement of Tully-Fisher relation and $M_r - v_{\mathrm{circ}}$ distribution best suited for comparison with such theoretical predictions.

\subsection{Motivation for this study}

Given the large body of literature on the TFR we need to justify revisiting the relation. The Calar Alto Legacy Integral Field Area provides us with three main reasons, two related to the observational data type at hand (Integral Field Spectroscopy -- IFS) and the last one tied to the available sample properties. 

First, optical long-slit observations have traditionally been the observational basis for the TFR analysis. 
Long-slit observations have the drawback of not being able to view the entire velocity field of a galaxy and thus being prone to being affected by non-circular velocity field distortions. IFS data allow to use the full velocity information available to correct for non-axisymmetric velocity field features, characterise the specific angular momentum of galaxies and distinguish between disturbed velocity fields/pristine disks/slow rotators. A possibility
to perform the sample selection using kinematic properties of galaxies is more relevant to the TFR than visual morphological classification. This is strikingly confirmed by \citet{Flores2006}. These authors show that the very large scatter in the intermediate redshift TFR, previously reported from  long or multi-slit spectroscopic observations, is a result of modelling a large fraction ($\approx$65\%) of galaxies with anomalous kinematics. A similar point is made by \citet{Andersen2003}, where the authors demonstrate that galaxies with large kinematic and photometric asymmetries in their velocity fields tend to be offset from the TFR. Using 2D velocity fields also prevents slit misalignment with the semi-major axis of the galaxy, even if such differences should not be a major problem in the TFR context \citep{Amram1994, Courteau1997, Giovanelli1997, Ballesteros2014}.

In addition, the observed line-of-sight rotation velocities must be de-projected in order to obtain the true circular rotation velocities. However, inclination is a notoriously difficult parameter to measure, and frequently the largest source of uncertainty in circular velocity measurement \citep{Schommer1993, Garrido2004, Giovanelli1997, Obreschkow2013}. In most of the Tully-Fisher studies inclination estimates are obtained from the variously defined apparent axis ratios $b/a$ in the following way \citep{Hubble1926}:

\begin{equation}
cos(i)^2 = \frac{(\frac{b}{a})^2 - q^2}{1 - q^2}
\label{eq:Hubble1926}
\end{equation}

Here $q$ is the intrinsic axis ratio of the galaxy, which is different for different galaxy morphological types  \citep[e.g. ][]{rodriguez2013intrinsic}, but a mean value of $q$ = 0.2 is frequently used \citep{Tully2000}. However, there are several shortcomings of this method. It frequently overestimates the inclination for face-on galaxies, because any irregularity (such as the spiral arms, bars, disk asymmetries) at the outskirts of a galaxy will make the $b/a$ seem higher \citep[e.g. ][]{maller2009intrinsic}. For galaxies that are close to edge-on, photometric inclination estimates suffer from uncertainty in the intrinsic axis ratio $q$, which depends on the Hubble type of the galaxy and evolves through cosmic time \citep{Obreschkow2013}. However, it is not
thought to be very significant in the context of TFR \citep{Courteau1997, Hall2012}.

We are able to circumvent the aforementioned difficulties with assumptions made for photometric inclination estimates
and directly model the velocity fields as rotating disks, obtaining kinematic inclination and position angles as free parameters. Such methods were first used a few decades ago by radio \citep{Rogstad1976, Bosma1978} and Fabry-Perot interferometry observers \citep{Marcelin1982, Bland1987, Nicholson1992, Schommer1993, Amram1994}.

To our knowledge, the first analysis of TFR using 2D IFS velocity fields was \citet{Courteau2003} study of barred
and unbarred disk galaxies. They used the SparsePak IFU \citep{Bershady2004} to test if the rotation velocities measured using long-slit spectroscopy are reliable. \citet{Andersen2003} derived a "face-on" Tully-Fisher relation based on 24 H$\alpha$ velocity fields of low-inclination (16\textdegree - 41\textdegree) galaxies. They show that kinematic inclination estimates are sufficiently accurate down to $\approx$15\textdegree, and that such an approach allows avoiding systematic and random errors arising from use of photometric axis ratio-based inclination estimates. For example, \citet{Schommer1993} find that photometric inclination estimates are systematically larger than inclinations derived from kinematics for galaxies with inclinations of up to 50\textdegree. 
As shown in this paper, by combining the photometric and kinematic data we can model the velocity fields consistently and obtain the full distributions of parameter uncertainties for inclinations, position angles and kinematic parameters, including the estimated rotation velocity.

Although several other IFS-based Tully-Fisher studies exist, they tend to focus on higher redshifts and have small sample sizes, aiming to investigate the dynamical state of high redshift galaxies, assembly times of rotating disks and morphological evolution \citep{Swinbank2006, Puech2008, Cresci2009, Gnerucci2011}. Similarly, \citet{Green2014} use the TFR measured from H$\alpha$ velocity fields of local gas-rich galaxies to gain insight into their high redshift analogues. Recently \citet{KMOS2016} use a sample of 18 $z ~ 1$  KMOS galaxies to demonstrate that the TFR obtained from H$\alpha$ emission is identical to the present one.

Therefore, the third major reason to revisit the TFR is the fact that CALIFA survey has well-defined sample selection criteria compared to earlier published work. Indeed, in virtually all past TFR studies, the authors aim for a "clean sample", meaning a set of late (usually Sa or later) morphological type galaxies with ordered circular motions \citep[e.g.][]{Courteau1996, Tully2000, McGaugh2000}. The goal of the majority of such studies was to estimate or calibrate the 'template' Tully-Fisher relation, best suited for distance measurements. Many methods to account for selection effects in TFR samples have been developed, including galaxy cluster observations to obtain volume-complete samples, the so-called inverse fitting, performing corrections based on morphology, extinction estimates, attempts to account for varying distances to cluster center, implicit sample incompleteness and many other possible sources of bias (see \citet{Giovanelli1997, Tully2000, Verheijen2001, Masters2006, Saintonge2011} and references therein for in-depth discussions of these methods and their shortcomings). While the above approaches are justified when the goal is to obtain a tight, linear distribution with the minimal amount of scatter, our objective is different. We aim to use data-driven modelling to obtain a volume-corrected 2D distribution of rotationally supported galaxies. While such a distribution would not be directly useful for distance measurements (although we provide the parameters of a standard linear fit) it could place more stringent constraints on galaxy evolution models than the standard TFR or luminosity and velocity functions separately.

A sample selected for such a purpose should span the widest possible range of morphologies \citep{Verheijen2001}. The wish to include galaxies of all types and be able to perform consistent volume and large-scale structure corrections has compelled us to use the stellar velocity fields in this study. Stellar velocity fields have rarely been used for TFR measurements, mainly due to their lower signal-to-noise level. Emission lines are more easily detected in spectroscopic data, meaning that velocity fields based on emission line kinematics will extend further out in the disk and have better spatial resolution. However, the main advantage of stellar velocity fields is that they can be obtained for galaxies without significant gas emission lines, i.e. early type galaxies. A significant fraction of early type galaxies follows the TFR: for example, \citet{Krajnovic2008} state that about 80\% of early type galaxies and S0s have a rotating disk component. Similarly, \citet{Emsellem2011} show that the majority of early type galaxies in the \textsc{ATLAS3D} sample are fast rotators, while \citet{Davis2011} demonstrate that early type galaxies lie on the CO Tully-Fisher relation, albeit likely offset from the one derived for spirals.

CALIFA is the first IFS survey to include many late type galaxies as well, to have a large and statistically well-defined sample and a sufficiently large field of view, all necessary for this sort of analysis. 

\section{Data, sample selection and characterisation}
\label{sec:sample}

\subsection{The CALIFA Survey}
CALIFA \citep[hereafter W14]{Sanchez2012, SCpaper} is a legacy IFS survey of 600 nearby galaxies. Observations use the PMAS instrument \citep{Roth2005} in PPaK \citep{Verheijen2004} mode,  mounted on the 3.5 m telescope at the Calar Alto observatory. The sample of galaxies being observed with CALIFA is drawn from a larger pool of galaxies, selected from the SDSS DR7 survey, and termed the "mother sample" (MS). The MS is primarily diameter-limited, aimed at using the detector area efficiently. The CALIFA selection criteria are described in \citetalias{SCpaper} in more detail. One salient feature of the CALIFA MS is that its selection criteria are well understood, providing us with a representative sample of galaxies that can be corrected for selection effects down to an absolute $r$-band magnitude of $-19$ mag. By adopting a probabilistic approach to outlier rejection we will show in this paper that we can keep a similar property of the sample even as we restrict it to conform to more stringent criteria. 

The CALIFA data have been reduced using the CALIFA pipeline and we refer to \citet{Sanchez2012, Husemann2013, Garcia-Benito2015} for all the details. The result of the data reduction is two spectral cubes of the target galaxy, one in the V1200 grating and one in the V500 grating, which can be used to extract kinematic information. V1200 grating has been used here because this setup allowed measuring the velocity dispersion down to $\approx$50 km/s. 

Kinematic information of the stars was extracted from CALIFA datacubes using the pPXF fitting procedure \citep{PPXF} and INDO-US \citep{indous2004} spectral templates library. Bad pixels, foreground contamination, low-quality spaxels with S/N < 3, emission line regions in the spectra and regions outside 3750–4650 \AA\ range (for the V1200 setup) were excluded from the fit. Spatial Voronoi binning \citep{cappellari2003adaptive} was applied to ensure a constant signal-to-noise ratio S/N = 20 in velocity dispersion and led to variable size bins with diameters ranging from 0.1 kpc to 21 kpc in linear size. On average the galaxies have 131 useful kinematic data points, with the lowest number being 1 (galaxies with too few Voronoi bins were excluded from further analysis, as described in the next paragraphs) and the highest number of bins being 760. The velocity and velocity dispersion values for each bin and the associated uncertainties were derived using 100 MCMC realisations of the fit. We refer the reader to the first paper of the CALIFA stellar kinematics series (Falc{\'o}n-Barroso et al., submitted) where the kinematic map extraction is described in full detail.

\begin{figure}[htbp]
\includegraphics[width=\linewidth]{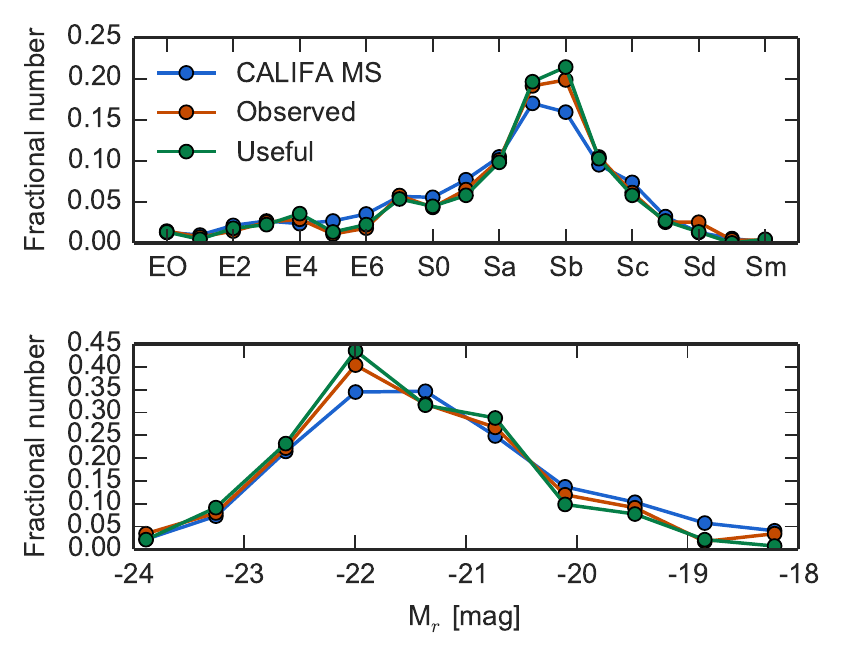} 
\caption{Normalised histograms of SDSS Petrosian M$_r$ and morphological types of the CALIFA mother sample, observed sample (as of October 2013) and the useful subsample that we included in our analysis.}
\label{fig:MS_samples}
\end{figure}



We started with the sample of galaxies observed by CALIFA until October, 2013. This corresponded to 277 objects with derived stellar velocity fields. We term this the "observed sample". Out of these, 51 were rejected at the beginning as non-usable. A significant fraction of the rejected galaxies, 17, had too few bins ($N_{bins}$ < 6) to even try to constrain the rotation curves with the stellar velocity fields reliably. In addition, we excluded 31 galaxies whose velocity fields could not provide a physical model. This included several heavily masked mergers, 4 galaxies which had foreground objects obscuring a significant part of the velocity map, 2 galaxies with significant dust lanes that obscured velocity fields, and the 5 galaxies excluded due to their unsuitability for volume correction procedure described below. We emphasize that the rejection here was not directly related to the internal properties of the galaxies that were relevant to this analysis (absolute magnitude, stellar mass, rotation velocity), but to problematic observational data that precluded making realistic models. The resulting sample contains 226 useful velocity fields, and we refer to it as the "useful sample" in the following sections. 

Fig. \ref{fig:MS_samples} shows normalised SDSS Petrosian absolute magnitude $M_r$ and morphological type histograms for the CALIFA mother sample, the observed sample and the useful sample. As shown in the top panel, the observed sample contains a slightly larger fraction of Sa-Sb type galaxies as compared to the CALIFA mother sample. Even though the observing selection should be random, observing constraints and spatial variation within the sample volume might have introduced this discrepancy.

The lower panel of Fig. \ref{fig:MS_samples} shows normalised histograms of $M_r$. The observed and especially the useful sample miss the least luminous galaxies as compared to the mother sample. This effect in the useful sample is exacerbated by the fact that we were more likely to reject intrinsically fainter, later type galaxies at this step, because these galaxies were more likely to have a lower number of Voronoi bins. This must affect the outcome of the volume correction we will perform at later steps. 

\subsection{Volume corrections}

The CALIFA sample is limited by two main selection criteria, including all galaxies within the SDSS DR7 footprint that have (i) redshifts within $0.005 < z < 0.03$, and (ii) isophotal angular extents within $40\arcsec < \theta < 79\farcs2$. This construction principle allows us to perform volume corrections using the $V_{\mathrm{max}}$ method \citep{Schmidt1968}, in much the same way as with a flux-limited sample (see W14 for details). While the sample as a whole is not volume-complete, each galaxy can be assigned a well-defined accessible survey volume $V_{\mathrm{max}}$ over which it \emph{would} be included given its properties and given the sample selection criteria.

It is important to realise that the selection by \emph{apparent} diameter in CALIFA does by no means introduce a bias in terms of \emph{linear} sizes of the galaxies in the sample, because of the broad redshift range. Within the ``completeness range'' of the sample ($-19 > M_r > -23.1$; see \citetalias{SCpaper}), low-luminosity and small galaxies match the angular diameter criterion close to the low redshift limit, while more luminous and larger galaxies occupy higher redshifts. By adding the contributions $1/V_{\mathrm{max}}$ of all galaxies in suitable bins, we can calculate a volume-corrected estimate of a distribution function of the galaxy population.  \citetalias{SCpaper} demonstrated that the galaxy luminosity function as well as the size distribution function estimated this way from the CALIFA mother sample are in excellent agreement with results from SDSS.

Another potential source of bias is related not to the sample selection process, but to the properties of the particular cosmological volume a given survey is probing. The CALIFA survey samples two nearby clusters (Virgo and Coma) as well as the underdensities in between, resulting in significant radial number density variations. \citetalias{SCpaper} showed how these radial variations can be quantified and absorbed into ``effective volume'' correction factors. All volume-corrected quantities shown in this paper use these effective $V_{\mathrm{max}}$ values. 

\begin{figure}[htbp!]
\includegraphics[width=\linewidth]{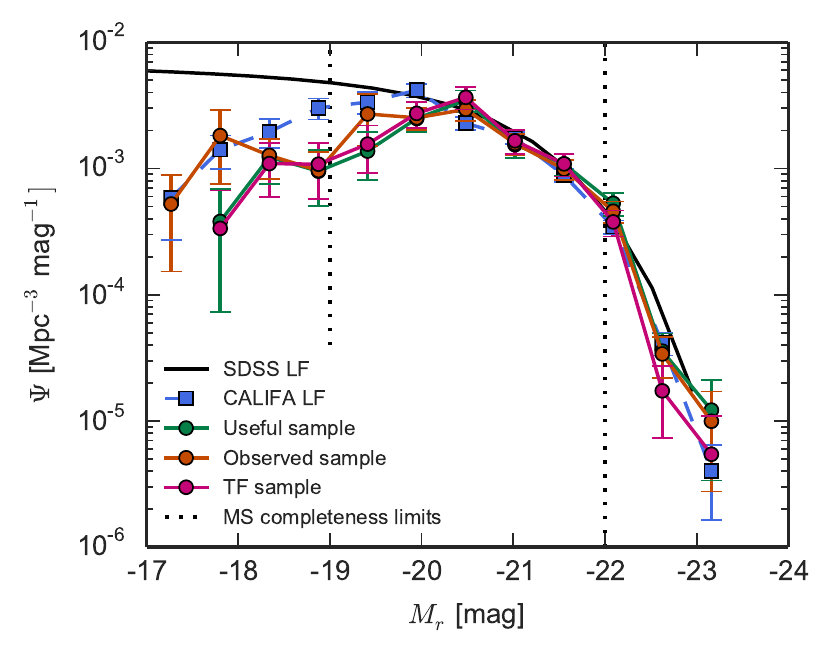} 
\caption{Luminosity functions of the CALIFA mother sample (MS), the observed sample, the useful sample (see text) and the final Tully-Fisher sample defined in Sec. \ref{sec:MoG}. The dotted lines denote the MS completeness limits calculated in \citetalias{SCpaper}. SDSS DR7 Petrosian $r$-band magnitudes were used for comparison with the SDSS luminosity function. The effects of outlier rejection steps are evident, especially for the more numerous low luminosity galaxies with higher $1/V_{\mathrm{max}}$ weights.}
\label{fig:LF_final}
\end{figure}

In this paper we further exploit the concept of volume corrections for the Tully-Fisher relation. Since our ``useful sample'' is much smaller than the CALIFA MS, the volume correction factors need to be adjusted to reflect the size of the subsample. As long as the sample is a random subset of the mother sample, it is sufficient to reduce $V_{\mathrm{max}}$ by the sampling rate; this concept was used in the first two CALIFA data releases to verify that the releases subsets (of 100 and 200 galaxies, respectively) are consistent in their statistical distribution properties with the mother sample, and with the galaxy population as a whole \citep{Husemann2013, Garcia-Benito2015}. In Fig.\ref{fig:LF_final} we used the same approach to compare the galaxy luminosity function constructed from our ``useful sample'' with the results from the CALIFA MS and from SDSS. The agreement is excellent for absolute magnitudes $M_r < -20$ mag, but there seem to be too few galaxies in the bins fainter than $-20$ mag for the ``useful sample''. As mentioned above, this is probably due to our rejection of some late-type galaxies from the kinematic analysis due to an insufficient number of Voronoi elements. Apart from this caveat we conclude that our sample can be seen as volume-representative for the local galaxy population, at least for absolute magnitudes $-22 < M_r < -20$.

Five more galaxies were excluded from further analysis: NGC 4676B and NGC5947 had been added to the CALIFA MS by hand (see \citetalias{SCpaper}) and have no associated $V_{\mathrm{max}}$ values, while for NGC 7625, NGC 1056, NGC 3057 these values are disproportionally small ($< 10^4$~Mpc$^{3}$). Since the $1/V_{\mathrm{max}}$ weights are then correspondingly large, these few galaxies would totally dominate any $1/V_{\mathrm{max}}$-weighted fit to the Tully-Fisher relation. Note that this is an inherent weakness of the $V_{\mathrm{max}}$  method. While more sophisticated approaches are conceivable that are more robust against such statistical fluctuations, we simply decided to remove these three objects. The effects of using or omitting  $1/V_{\mathrm{max}}$ weights when fitting the TFR are shown in Sec. \ref{sec:TF}.

\begin{figure}[htbp]
\includegraphics[width=\linewidth]{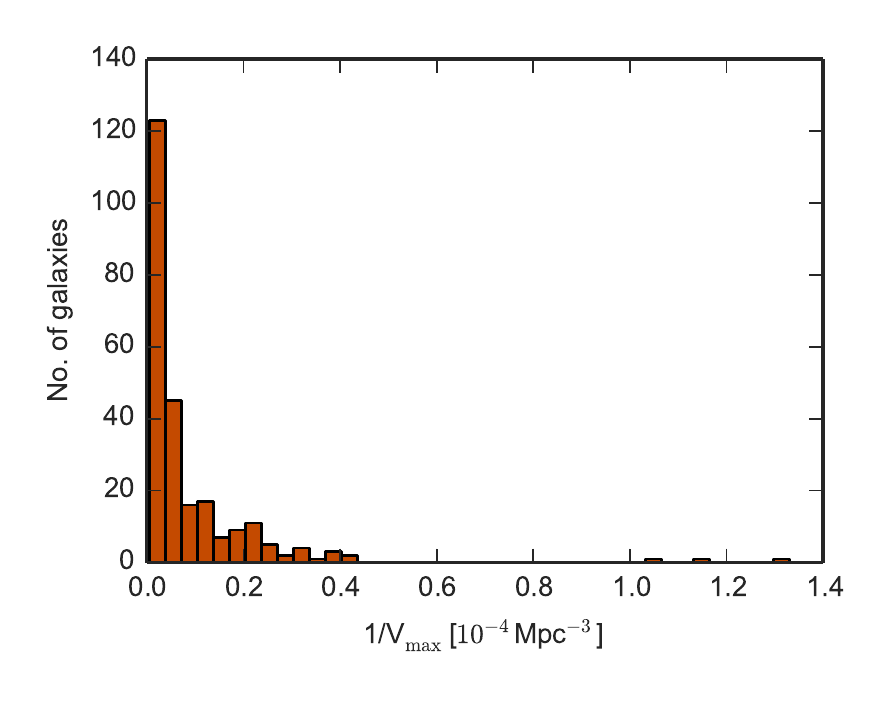} 
\caption{Volume and radial-density weight factor $1/V_{\mathrm{max}}$ histogram. The three outliers with $1/V_{\mathrm{max}}$  > 0.0001 were excluded from further analysis.}
\label{fig:Vmax_hist}
\end{figure}

\section{Luminosity data}
\label{sec:lumdata}

\subsection{Observed magnitudes}

We used the $r$-band growth curve photometry measurements described in \citetalias{SCpaper}. The uncertainties provided account for the combination of the contributions from the dark current and read noise, Poissonian sky counts error, uncertainties due to sky subtraction and an estimate of uncertainties arising due to masked foreground objects. The formal errors due to shot noise make only a small contribution to the error budget due to the large apparent sizes of our galaxies on the SDSS images. The magnitudes are corrected for Galactic extinction using SDSS pipeline values.

Absolute magnitudes were calculated using the prescriptions presented in \citetalias{SCpaper}. In short, the redshifts were corrected for Virgo-centric, Shapley and Great Attractor infall motions using the model by \citet{Mould2000}. K-corrections were determined from spectral energy distributions as described in \citetalias{SCpaper} and \citet{Walcher2008}. The distance uncertainties were derived from group velocity dispersions obtained by the collaboration in \citetalias{SCpaper} and combined with the photometric uncertainties when calculating the absolute magnitude uncertainties.

\subsection{Corrections for intrinsic absorption} 

The absolute magnitudes should be corrected for internal extinction, which depends on the inclination, bandpass and morphology in a non-trivial way. We adopted the methods described in \citet{Wild2011, Wild2011optical}, which provides dust attenuation as a function of the photometric axis ratio $b/a$, specific star formation rate and presence or absence of a significant bulge. We used the star formation rates determined from CALIFA H$\alpha$ line emission based on the prescriptions of \citet{Calzetti2013} \citep{Catalan2015}, photometric stellar masses determined by the collaboration \citepalias{SCpaper} and integrated CALIFA H$\alpha$  and H$\beta$ fluxes. The
corrected magnitude $M_r^c$ was calculated as $M_r^c = M_r + \Delta M_r$.

In order to correct the magnitudes for intrinsic attenuation we needed the Balmer decrement of our sample galaxies. The emission line properties of 30" radius aperture spectra were extracted
from the V500-datacubes of the galaxies. This aperture is large enough to include
virtually 100\% of the FoV of the CALIFA datacubes, without the need
to select a different aperture for each galaxy.

To extract the information contained in the spectra, we followed the
procedures described in \citet{Sanchez2014}, using the fitting package
{\sc FIT3D}\footnote{\url{http://www.caha.es/sanchez/FIT3D/}}.

Individual emission line fluxes were measured using {\sc FIT3D} in the
stellar-population subtracted spectra performing a
multicomponent fitting using a single Gaussian function. 
By subtracting a
stellar continuum model derived with a set of SSP templates, we are correcting for the effect of underlying stellar absorption,
which is particularly important in Balmer lines (such as H$\beta$).

\begin{figure}
\includegraphics[width=\linewidth]{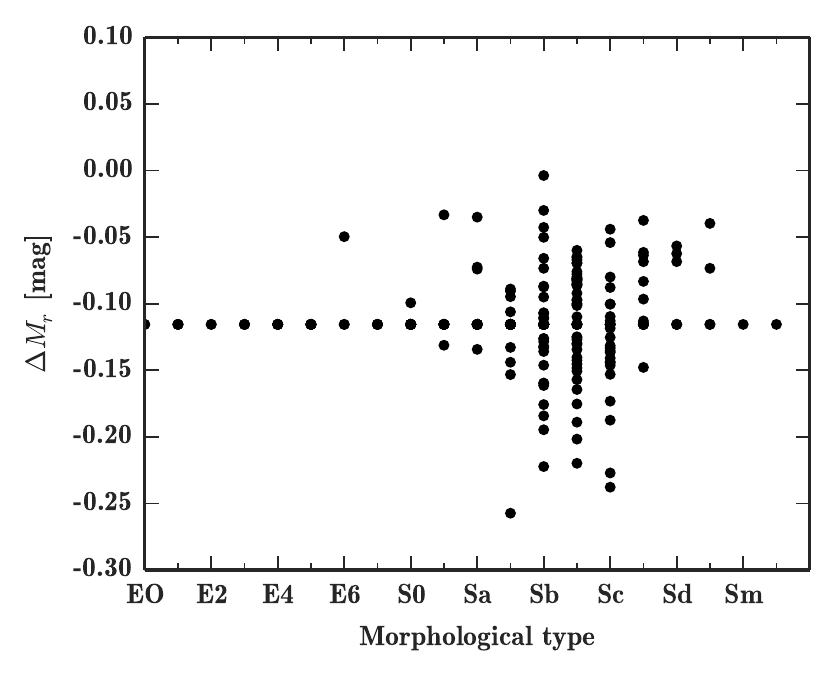}
\caption{The extinction correction vs. visual morphological classification.}
\label{fig:corr_morph} 
\end{figure}

\begin{figure}
\includegraphics[width=\linewidth]{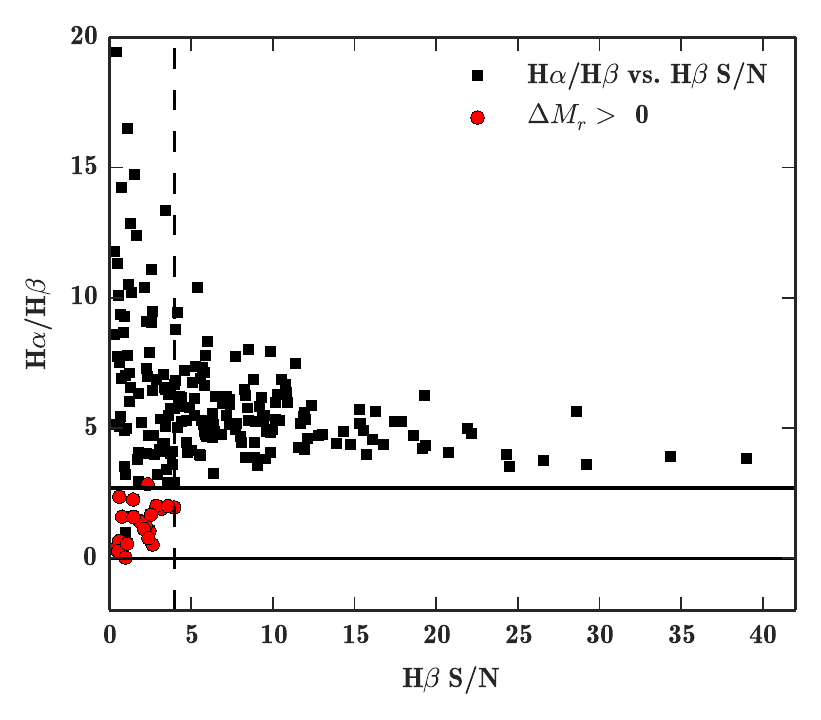}
\caption{Balmer decrement vs. $H_{\beta}$ signal-to-noise ratio. The red points are the galaxies that had unreliable extinction corrections due to their low S/N and correspondingly erroneous Balmer decrement values. The horizontal and vertical dashed lines show the region of likely untrustworthy extinction correction estimates. We used the mean magnitude of the extinction correction for these galaxies, see text.}
\label{fig:BalmerSNR}
\end{figure}

Visual morphological classifications were used to distinguish between galaxies with and without significant bulges which have different dust correction prescriptions. We assumed that galaxies that had been classified as Sc and later had no significant bulge. The correspondence between the visual morphological classification and the magnitude of the extinction correction is shown in Fig. \ref{fig:corr_morph}. 

It should be noted that we have used kinematic inclination values obtained in the following section elsewhere in the analysis, especially when correcting the rotation curves for inclination. However, the methods described in \citet{Wild2011, Wild2011optical} were derived using the photometric axis ratio as an input, therefore we employ it in our analysis for the sake of consistency.

We clipped the star formation rates and axis ratios to the maximum values provided in \citet{Wild2011} (0.3 \textless $b/a$ \textless\ 0.9, -10.2 \textless\ $\log(sSFR)$ \textless\ -9.3 yr$^{-1}$ for bulge-dominated galaxies and -10.0 \textless\ $\log(sSFR)$ \textless\ -9.1 yr$^{-1}$ for disk-dominated galaxies).

The CALIFA H$\alpha$  and H$\beta$ fluxes are not reliable at the low S/N limit, leading to unrealistic Balmer decrements. We settled on making a S/N cut at H${\alpha}$/H${\beta}$ = 2.7 (S/N was very close to 4 there, see Fig. \ref{fig:BalmerSNR} for an illustration). For galaxies below the S/N = 4 limit and for those that had no reliable H${\alpha}$ or H${\beta}$ fluxes we adopted the average extinction correction value $\Delta M_r= -0.11$ mag.

\section{Velocity field modelling}
\label{sec:velfields}

\subsection{Model description}
\label{sec:vel_models}
It was long noted that the deprojected rotation curves of galaxies show a variety of shapes \citep{Rubin1985, Verheijen2001}, exhibiting differences attributed to morphology \citep{Rubin1985, Verheijen2001} or luminosity \citep{Persic1991}. Several parameterisations of rotation curves exist, some being purely phenomenological \citep{Courteau1997, Vogt1996, Rix1997a}, some attempting physical parameterisation \citep{Persic1991, Bosch2000, Persic1996}.

We have attempted using a variant of the $arctan$ function \citep{Courteau1997}, the hyperbolic tangent \citep{Neumayer2011}:
\begin{equation}
v(r) = v_0 + \frac{2}{\pi} v_{c} \cdot tanh\left[\frac{r}{k\cdot r_{50}}\right]
\label{eq:tanhRC}
\end{equation}

\noindent where $v_0$ is the recession velocity, $v_c$ is a free parameter governing the amplitude of the rotation curve, $k$ describes the sharpness of RC turnover, and $r_{50}$ is the optical half-light semi-major axis, determined from $r$ band growth curve photometry. In addition, a galaxy is allowed to have arbitrary inclination and position angles. We do not allow the kinematic center position to vary, since the spatial resolution of Voronoi bins is variable and sometimes too low to provide a meaningful constraint on the center position. 

The main attraction of this model was its simplicity, i.e. the lowest number of free parameters. However, we found that this simple model could not fit the rising or falling rotation curves.

At the cost of parameter degeneracy we assumed another model for RC shapes, implemented in \citet{Bertola1991} and also discussed in \citet{Boehm2004}: 

\begin{equation}
v(r) = v_0  + \frac{v_{c}r}{(r^2 + k^2)^{\frac{\gamma}{2}}}
\label{rot_curve}
\end{equation}

The model has four free parameters -- $v_c, k, v_0$ and $\gamma$. The $v_c$ and $k$ parameters here take on similar roles as in the $tanh$ model (Eq. \ref{eq:tanhRC}), however, their values are different. The $\gamma$ parameter, usually varying between 0.8 and 1.2, allows modelling rising and falling rotation curves. A flat rotation curve is obtained when $\gamma$ = 1. 

In most optical studies the TFR defines the rotation velocity as measured at the outer part of a galaxy where the rotation curves are no longer rising. To achieve that we had to extrapolate the rotation curves of galaxies that were not sampled up to the point of turnover. 

We take advantage of an open-source Python implementation \citep{Foreman-Mackey2013} of affine-invariant MCMC
sampler \citep{Goodman2010}, called through a customised wrapper. MCMC methods provide the full posterior distributions of model parameters, leading to more realistic uncertainty estimates.

In order to constrain the physical parameters of the models, we applied a truncated Gaussian prior on the $\gamma$ parameter, effectively constraining it to lie between 0.8 - 1.2. We find that this range of $\gamma$ values describes the range of physically possible rotation curves well and helps avoid degeneracies. In addition, we
constrain the marginalised rotation velocity at 2.2 scale lengths ($v_{\mathrm{2.2}}$) to be below 600 km/s, and restrict $k$ > 0.

\begin{figure*}[hbtp]
\includegraphics[width=\linewidth]{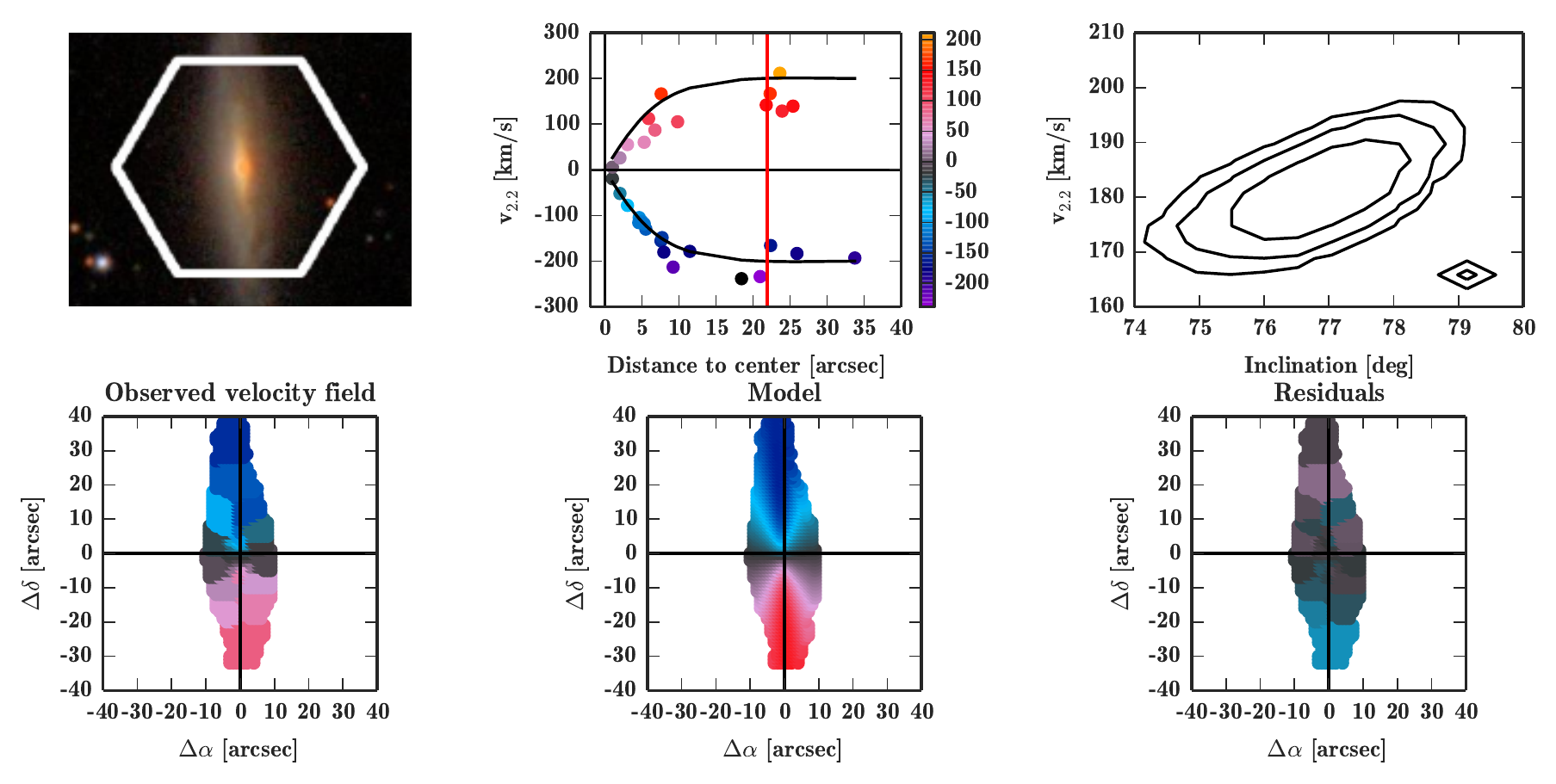}  
\caption{A velocity field of an inclined galaxy (IC5376, $i_{phot}$ = 80\textdegree). Top: SDSS composite image (left), model (black) and measured rotation curves (middle), joint $i$ - $v_{\mathrm{2.2}}$ distribution (shown as 1, 2, 3 standard deviation contours). Bottom: observed velocity field (left), model (middle) and residuals. The red line indicates the location of 2.2l$_{sc}$. The velocity scale is the same in all panels.}
\label{fig:low_bins}
\end{figure*}

Even in the case of the simplest hyperbolic tangent model we observed a strong coupling between the inclination angle and the $v_c$ parameter, i.e. the amplitude of the rotation curve. This is somewhat expected, as almost any velocity field can potentially be modelled as a fast-rotating, almost face-on disk, or an inclined one with a lower intrinsic rotation velocity if the spatial resolution is low. This proved to be a problem for highly-inclined galaxies with a small number of bins (Fig.\ref{fig:low_bins}), because the inclination could not be constrained well.

In order to break this degeneracy, we introduce a truncated Gaussian prior on the inclination angle for galaxies with $i_{phot} > 75^{\circ}$. We estimate the prior inclination using Eq. \ref{eq:Hubble1926} and the photometric axis ratios provided in \citetalias{SCpaper}, assuming the intrinsic disk thickness $q$ = 0.2 and the standard deviation of the Gaussian $\sigma$ = 3\textdegree.  

All photometric axis ratios were inspected visually and found to be quite accurate inclination indicators for highly-inclined galaxies. This is not necessarily the case for low-inclination galaxies that can look less circular than they are due to spiral arms and other irregularities.

We have noticed a strong bimodality in the marginalised $v_{\mathrm{2.2}}$ distribution for a number of objects. It occurred either due to a low number of Voronoi bins and therefore poorly constrained inclination, or the inability to constrain the model based on kinematic information alone. In such cases, an identical truncated Gaussian prior was placed to constrain the inclination (and, consequently, $v_{\mathrm{2.2}}$) to a more plausible range.

\begin{figure*}[hbtp]
\includegraphics[width=\linewidth]{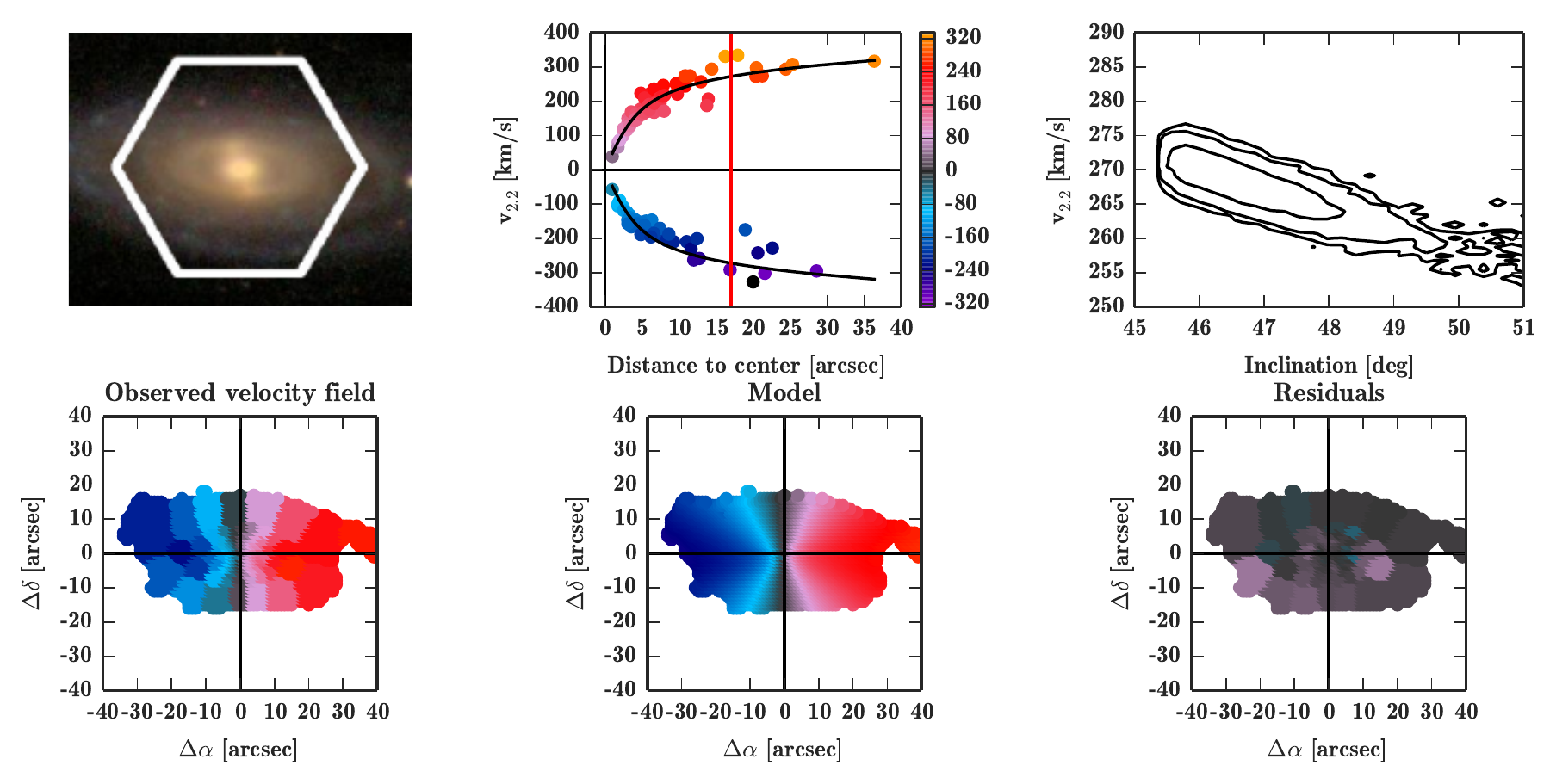}  
\caption{
Top: SDSS composite image of NGC1645 (left), model (black) and measured rotation curves (middle), joint $i$ - $v_{\mathrm{2.2}}$ distribution. Bottom: observed velocity field (left), model (middle) and residuals. The photometric inclination was estimated to be equal to 52\textdegree.}
\label{fig:vel_field_resid}
\end{figure*}

We take 480000 MCMC samples for each galaxy, rejecting the first 160000 to reduce the impact of the choice of initial parameters. The chain lengths were chosen after repeated modelling has shown that the models selected were robust, i.e. the parameter distributions did not change between fits. The MCMC outputs provide the full distributions of rotation curve parameters, inclination and position angle values for each galaxy. From these we can obtain the marginalised posterior distribution of the modelled velocity at a given radius, as well as kinematic inclination and position angle estimates. Two examples of observed and model velocity fields, as well as their rotation curves and joint inclination-$v_{\mathrm{2.2}}$ distributions are shown in Fig. \ref{fig:low_bins} and \ref{fig:vel_field_resid}. 

The mean values and standard deviations of kinematic inclination and position angles were determined through directional statistics, i.e. by calculating the vector means and circular standard deviations of the chain values. Although the inclination and position angles were allowed to vary freely during fitting, the resulting chain values were wrapped to intervals $[0; 90]$ and $[0; 180]$ respectively.

\subsection{Definition of rotation velocity measure}
When the rotation velocity is measured at the outer parts of a galaxy, the Tully-Fisher relation links the halo properties and the baryonic mass. However, in practice it is difficult to connect the true halo-induced velocity and the measured velocity due to limited radial coverage \citep{Verheijen2001}. The rotation velocity has to be measured at a particular point of the rotation curve, which affects the slope of the TFR \citep{Yegorova2007}. 

It is often measured at $r_{\mathrm{2.2}}$ = 2.2 $l_{\mathrm{sc}}$, where $l_{\mathrm{sc}}$ is the exponential scale radius of the disk, as well as at $r_{\mathrm{opt}}$, the radius containing 83\% of all light \citep{Courteau1997}. Other, non-parametric definitions are also employed, for example, the maximum rotation velocity $v_{\mathrm{max}}$, rotation velocity at the flat part of the rotation curve ($v_{\mathrm{flat}}$), the mean value of the outermost points of a rotation curve, etc (see \citet{Boehm2004} for a discussion).

Two practical measures are $v_{\mathrm{2.2}}$ and $v_{\mathrm{opt}}$. We were reluctant to use $v_{\mathrm{max}}$, $v_{\mathrm{flat}}$ and other non-parametric measures of the circular velocity because a) there often were outlier points on the rotation curves due to Voronoi binning or lack of masking, b) not all the rotation curves were asymptotically flat. We decided to use $v_{\mathrm{opt}}$ as our velocity measure, because it is straightforward to compare with simulations and other observations as opposed to $v_{\mathrm{2.2}}$ measuring which requires structural decomposition of a galaxy.

\begin{figure}[hbtp]
\includegraphics[width=\linewidth]{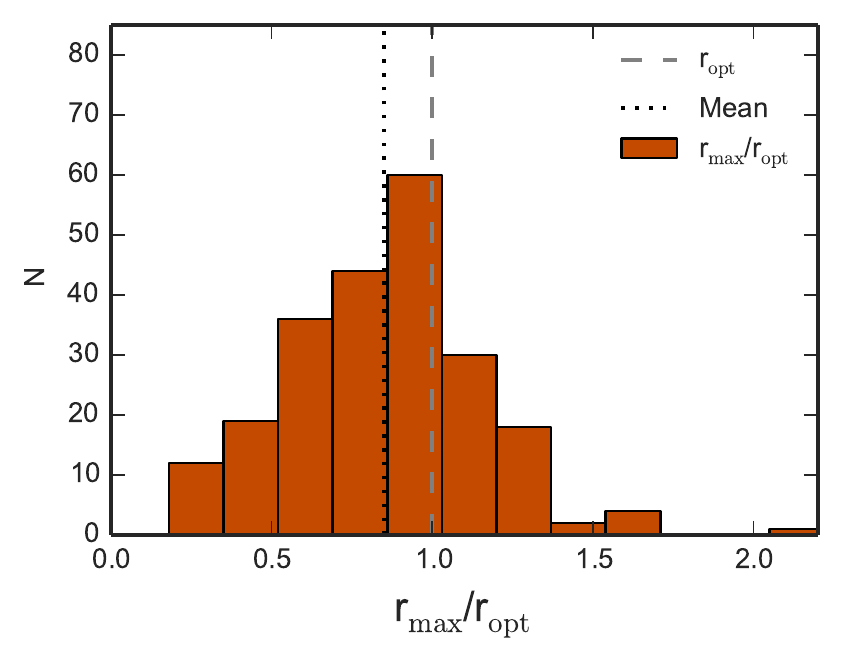}
\caption{Relative spatial coverage histogram of CALIFA stellar velocity fields. The dotted line denotes the mean value within the sample, the dashed one is at r$_{\mathrm{opt}}$.}
\label{fig:coverage}
\end{figure}

The mean coverage of CALIFA velocity fields, i.e. the maximum radius divided by $r_{\mathrm{opt}}$ is shown in Fig.\ref{fig:coverage}. 
 
In addition, we have checked if the galaxies with lower spatial coverage are offset from the Tully-Fisher relation. We do not find a significant offset (see Fig.\ref{fig:TFR} in  Sec.\ref{sec:TF}).

\begin{figure}[hbtp]
\includegraphics[width=\linewidth]{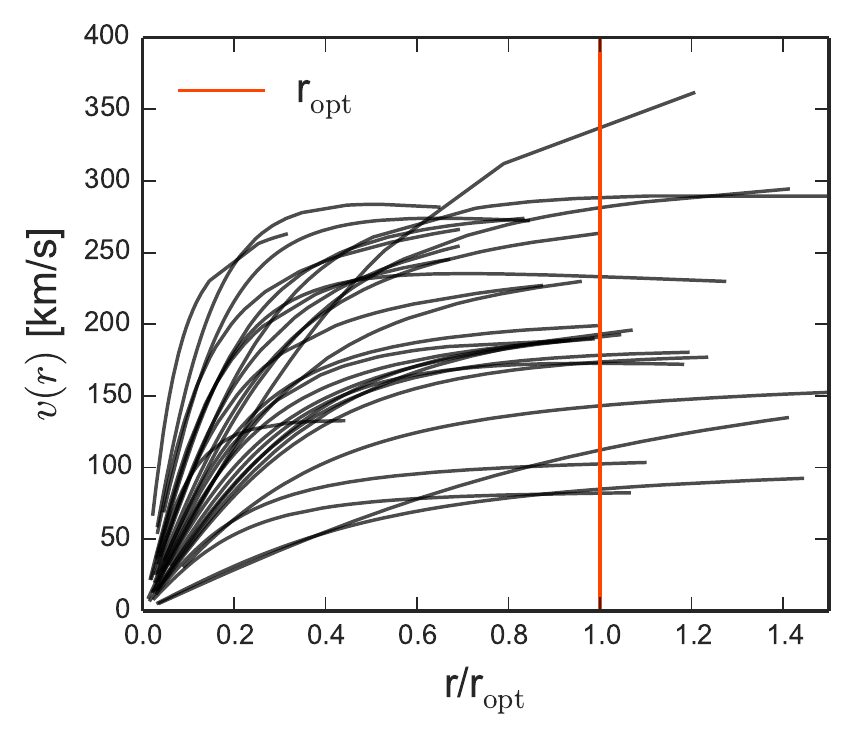} 
\caption{Random sampling of 30 rotation curves scaled to $r_{\mathrm{opt}}$, shown as the vertical line.}
\label{fig:radii_comparison}
\end{figure}

\subsection{Modelling results and uncertainties}
\label{sec:modelling_results}

For each galaxy we estimated the rotation velocity $v_{\mathrm{opt}}$ choosing between the two models described in \ref{sec:vel_models}. For the absolute majority (all except 9) of the galaxies the more complex \citet{Bertola1991} model was preferable. In the 3 cases when both models were clearly wrong in the outer parts of the galaxies we picked the mean value of the last two points and added 20 km/s to the rotation velocity uncertainty.

\begin{figure}[hbtp]
\includegraphics[width=\linewidth]{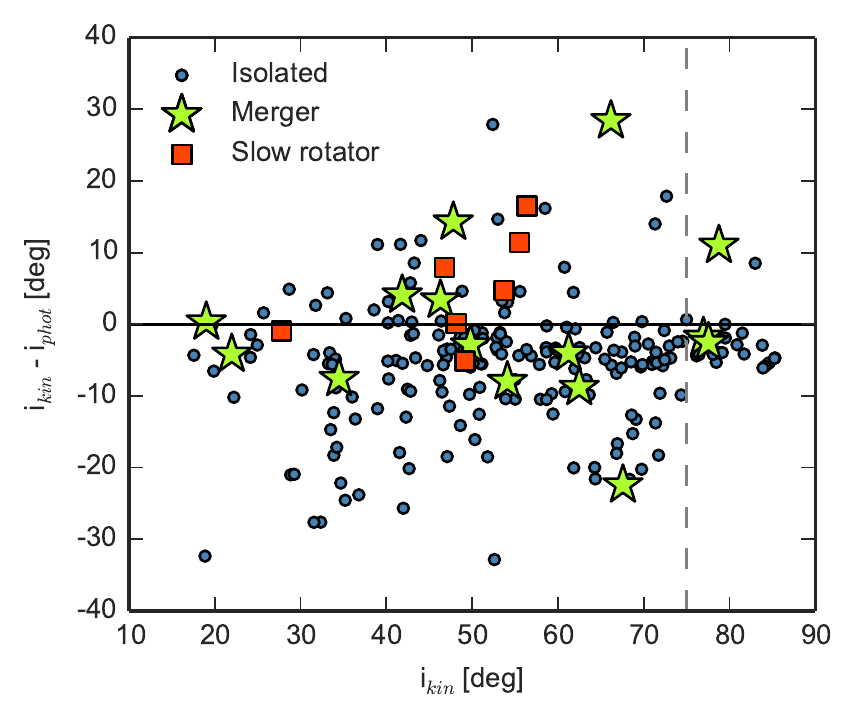}

\includegraphics[width=\linewidth]{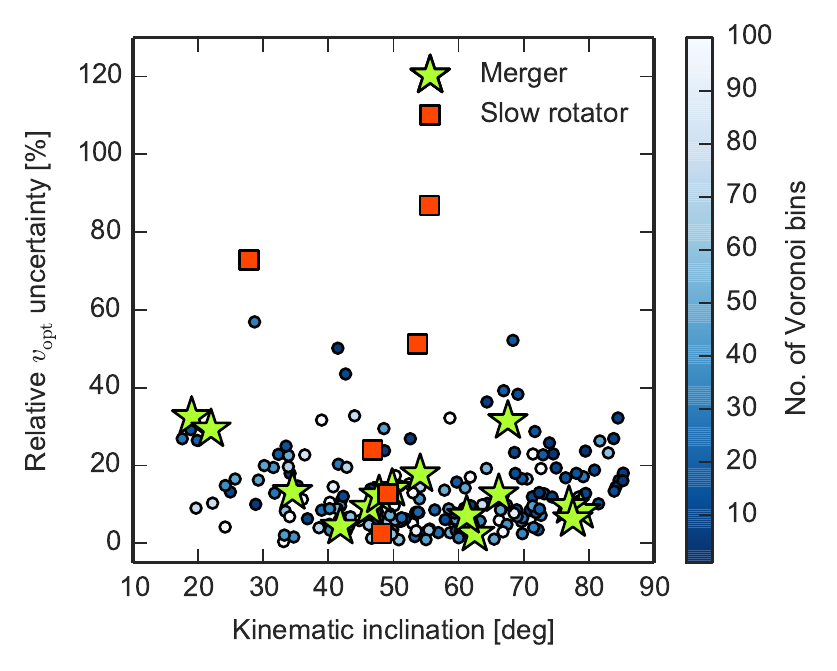}
\caption{Top: Difference between kinematic and photometric inclination estimates, calculated from $r$-band axis ratios assuming the intrinsic disk thickness $q$ = 0.2. Slow rotators and interacting galaxies are marked in red and light green respectively. Bottom -- $v_{\mathrm{opt}}$ uncertainties vs. kinematic inclinations.}
\label{fig:kin_phot_comparison}
\end{figure}

Fig.\ref{fig:kin_phot_comparison} shows the comparison of kinematic and photometric estimates of inclination. As would be expected \citep{Schommer1993}, photometric estimates for inclination are systematically higher for low-inclination galaxies, because any irregularity in the apparent light distribution forces the axis ratio towards lower values. At higher inclinations, inclination estimates of galaxies classified as mergers and slow rotators tended to differ the most. Nevertheless, this does not present a problem for TF studies because the intrinsic rotation velocity is obtained by dividing the line of sight rotation velocity by the sine of the inclination angle, and the slope of the sine function is shallow for angles above 75\textdegree. We employ an identical Gaussian prior as above for the majority of galaxies with $i \geq$ 75\textdegree , therefore kinematic and photometric estimates tend to converge at the highest inclinations.

In order to estimate the reliability of $v_{\mathrm{opt}}$ measurements, we discuss the potential sources of uncertainties. MCMC modeling provides estimates for uncertainties inherent in modelling a particular galaxy.

For some galaxies the main source of velocity uncertainties was the limitations of the simple rotating disk model. For instance, mergers and slow rotators showing little or no ordered rotation could not be well constrained and had broad, non-Gaussian posterior distributions of $v_{\mathrm{2.2}}$.  Relative $v_{\mathrm{opt}}$ uncertainties vs. kinematic inclinations are shown in Fig.\ref{fig:kin_phot_comparison}, demonstrating the difficulty in constraining the rotation velocity of slow rotators. Due to their large velocity uncertainties, they would not contribute significantly to the fit of the Tully-Fisher relation. However, the majority of such galaxies were rejected from the final Tully-Fisher sample as described in Sec.\ref{sec:MoG}.

Other potential sources of uncertainties include the limited CALIFA field of view and the spatial resolution of the binned stellar velocity fields. However, they are either implicitly included in the posterior $v_{\mathrm{opt}}$ uncertainties or too difficult to estimate within the scope of this paper. A careful analysis of uncertainties in the template TFR context is presented in \citet{Saintonge2011}.

In most previous TF work the photometric axis ratio $b/a$ has been directly converted into inclination, without assuming any associated uncertainties or potential difference between kinematic inclination (i.e. the real inclination of the observed rotating component) and its photometric estimate based on $b/a$. This has led to rotation velocity estimates having uncertainties of the order of few km/s, and Tully-Fisher relations with negligible uncertainties on the line fit parameters. In our opinion, using kinematic inclination values is more justified than deriving them independently from the photometric axis ratio. Also, MCMC modelling of the velocity fields provides the full posterior distribution of velocity uncertainties. Such consistent, self-contained rotation curve modelling is only possible with IFS data. 

However, the stellar velocity fields we are using here have their own share of problems, such as large Voronoi bins and limited spatial extent, and therefore likely to lead to larger velocity measurement uncertainties than the other methods. A direct comparison on the velocity uncertainties derived using the different methods and data (such as stellar/gas 2D velocity fields or long slit observations) is outside the scope of this paper. We just state that the consistent velocity uncertainties provided by the methods presented here can be propagated into Tully-Fisher relation fit, providing more reliable constraints on its internal scatter. 

\subsection{Calculating the circular velocity}
\label{ADC}

We are using stars as the tracer of the circular velocity of the galaxies. Stellar velocity fields have the advantage of being available for all morphological types and also of being less distorted than gas fields \citep{Adams2012, Kalinova2015}. However, stars are dynamically hot tracers and a so-called asymmetric drift correction, which takes the velocity dispersion into account, is frequently applied in order to obtain the circular velocity $v_{\mathrm{circ}}$. 

According to \citet{Kalinova2015}, the 'classic' \citep{Weijmans2008} asymmetric drift correction (ADC) underestimates the real underlying potential if the local inclination-corrected rotation velocity and velocity dispersion ratio $V/\sigma$ is smaller than 1.5. This implies that asymmetric drift corrections would not be accurate for the majority of galaxies within our sample. Even though the asymmetric drift correction changes the shape of the inner rotation curve dramatically, the estimated circular velocity does not change significantly for rotation-supported galaxies. However, the classical ADC approach is unsuitable when the assumption of a thin disk is not valid.

\begin{figure}[hbtp]
\includegraphics[width=\linewidth]{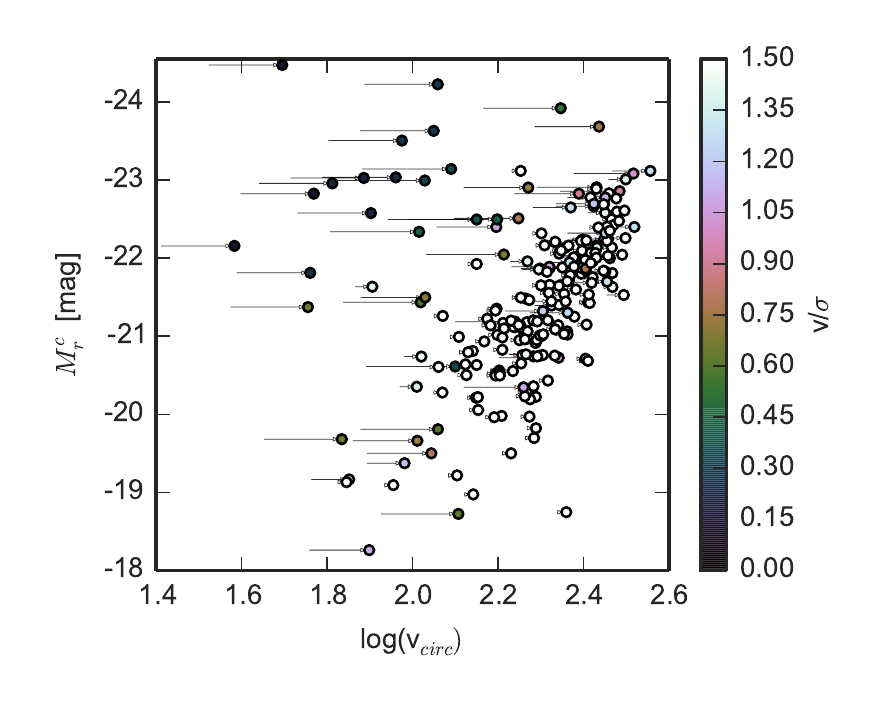}

\caption{Magnitude of circular velocity correction is shown as arrows. The points are the final circular velocity values, color-coded for the $v_{\mathrm{opt}}$/$\sigma$ factor. The galaxy with the lowest $\log(v_{\mathrm{circ}})$ value, a slow rotator, is not shown for clarity. This galaxy, NGC 6515 ($\log(v_{\mathrm{circ}})$ = 1.05), was excluded from further analysis during the outlier rejection procedure described in Sec. \ref{sec:MoG}.}
\label{fig:circ_vel_arrows}
\end{figure}

We decided to avoid the classical ADC in order to treat our sample in a consistent, homogeneous manner. Although advanced dynamical modelling is outside the scope of this observational paper, we decided to apply an empirical correction based on the findings of \citet{Kalinova2015}. They analyse the difference between dynamical masses inferred using classical ADC models and axisymmetric Jeans anisotropic Multi-Gaussian (JAM) models applied to stellar mean velocity and velocity dispersion fields of 18 late type galaxies observed with SAURON IFS instrument. We use the relation derived from Table 4 of \citet{Kalinova2015} and calculate the circular velocities by multiplying the measured velocity by the square root of the factors provided, based on the local measured $v_{\mathrm{opt}}$/LOS\ $\sigma_{\mathrm{opt}}$ of a galaxy. The uncertainty of the calculated circular velocity was calculated by adding in quadrature 10 km/s multiplied by 1 + $\Delta V$ (where $\Delta V$ is the square root of the uncertainty factor from \citet{Kalinova2015}) to the $v_{\mathrm{circ}}$ uncertainty budget. The magnitude of the correction is shown in Fig. \ref{fig:circ_vel_arrows}.

We have compared the gas rotation velocities of 44 galaxies obtained from CALIFA DR1 data \citep{Garcia-Lorenzo2015} with their stellar circular velocity values, shown in Fig.\ref{fig:circ_vel_gas_vopt_comparison} below. Ionised gas rotation curves were obtained from the envelope of the position-velocity diagram and corrected for inclination using photometric axis ratio $b/a$, then gas $v_{\mathrm{opt}}$ values were evaluated at the optical radius.

This is not a direct and accurate comparison, for many reasons. First of all, even though ionised gas is typically less dynamically hot than the stars, the measured gas rotation velocity is not tracing the gravitational potential directly and the gas dispersion needs to be taken into account. This presents additional difficulties because the gas dispersion cannot be measured directly with CALIFA spectral resolution. Furthermore, thermal motion and gas turbulence also contribute to the total gas velocity dispersion and cannot be distinguished from gravitationally induced velocity dispersion without additional data \citep{Weijmans2008}. In addition, the gas $v_{\mathrm{opt}}$ was estimated differently than the stellar rotation velocity. Photometric inclination estimates were used instead of kinematic ones, also no rotation curve modelling was performed, therefore any warps or distortions present in ionised gas were not accounted for. The most noticeable outliers in Fig.\ref{fig:circ_vel_gas_vopt_comparison} are offset due to the latter reasons. Despite this, for the majority of galaxies the two quantities are close to each other, with the stellar v$_{\mathrm{circ}}$ being typically larger as expected from the arguments above. A similar comparison for several CALIFA galaxies, using a different asymmetric drift correction method, is shown in \citet{Aguerri2015}.

\begin{figure}[hbtp]
\includegraphics[width=\linewidth]{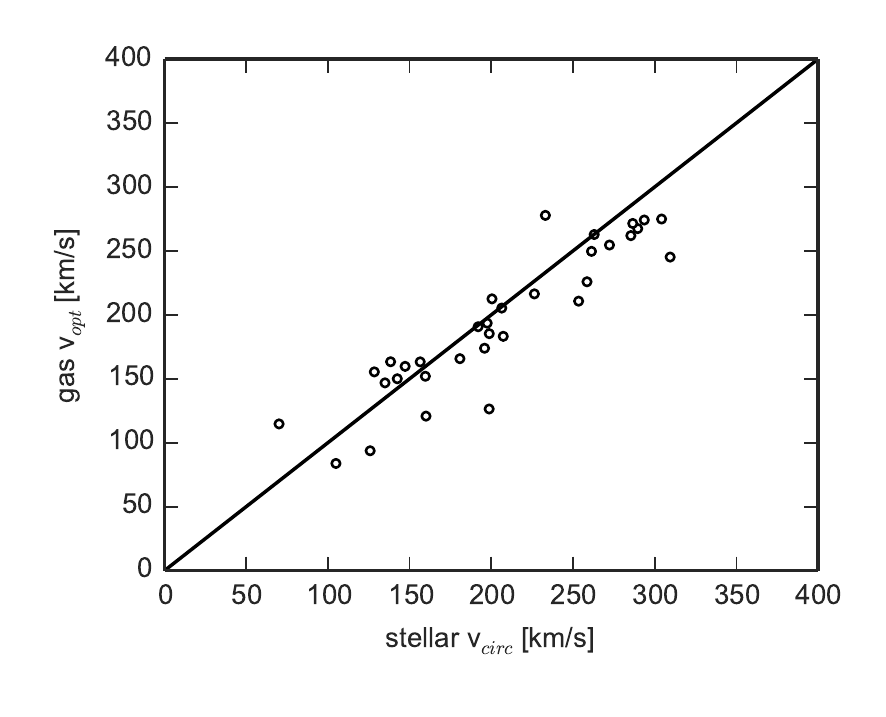}
\caption{Comparison between ionised gas rotation velocity at the optical radius and the stellar circular rotation velocity for 44 CALIFA DR1 galaxies.}
\label{fig:circ_vel_gas_vopt_comparison}
\end{figure}

We shall use the calculated circular velocity values in all further analysis, unless noted otherwise.

\section{Separation of different populations of galaxies in the M$_r$ - $v_{\mathrm{circ}}$ plane}

\subsection{Specific angular momentum}

The specific angular momentum $j$ and the total mass are key properties of galaxies that strongly influence their morphology, luminosity and secular evolution. A directly measurable quantity in IFS observations, related to $j$, is the $\lambda_{R}$ parameter, defined in \citet{Emsellem2007} as

\begin{equation}
\label{eq:lambda_par}
\lambda_R \equiv \frac{\langle R \, \left| V \right| \rangle }{\langle R \, \sqrt{V^2 + \sigma^2} \rangle}
\end{equation}

In practice $\lambda_{R}$ is calculated in the following way \citep{Emsellem2007}:

\begin{equation}
\lambda_R = \frac{\sum_{i=1}^{N_p} F_i R_i \left| V_i \right|}{\sum_{i=1}^{N_p} F_i R_i \sqrt{V_i^2+\sigma_i^2}} \, ,
\label{eq:lambda_IFU}
\end{equation}

\noindent where F$_i$, R$_i$, V$_i$ and $\sigma_j$ are the fluxes, semi-major axis values, velocities and velocity dispersion values of a spatial Voronoi bin $i$.


Measurements of $\lambda_{R}$ parameter values are available for the galaxies in our sample from work done within the CALIFA team. For CALIFA galaxies, the $\lambda_{\mathrm{Re}}$ parameter ($\lambda_R$ within 1 effective radius $R_e$) was calculated as described by Eq. \ref{eq:lambda_IFU} and corrected for inclination as described in Falc{\'o}n-Barroso et al. (in prep.), also see \citet{Barroso2015} and \citet{Querejeta2015}. Briefly, ellipticities $\epsilon$ were obtained from {\sc IRAF} ellipse fit models of the SDSS $r$-band images and the probability of observing a galaxy with an inclination $i$, given its ellipticity $\epsilon$, was calculated as

\begin{equation}
f(i|\epsilon) = \frac{f(q)(1 - \epsilon)}{\sqrt{sin^{2}i - \epsilon(2 - \epsilon)}}
\end{equation}

\noindent where $f(q)$ is the intrinsic shape distribution of galaxies. The $\lambda_{\mathrm{Re}}$ values were available for 206 out of 226 galaxies, because the authors rejected the galaxies with a low number of bins (typicaly $\leq$ 10) and interacting galaxies showing obvious kinematic irregularities from their calculation.

Although minor inconsistencies arise due to the use of slightly different parameters (such as inclinations) in this analysis, we emphasize that the lambda parameter is used in this study only as a qualitative illustration of the degree of rotation support in our sample galaxies. These minor inconsistencies and lack of $\lambda_{\mathrm{Re}}$ values for some of the galaxies thus have no influence on any quantitative result in the paper. 

\begin{figure}[hbtp]
\includegraphics[width=\linewidth]{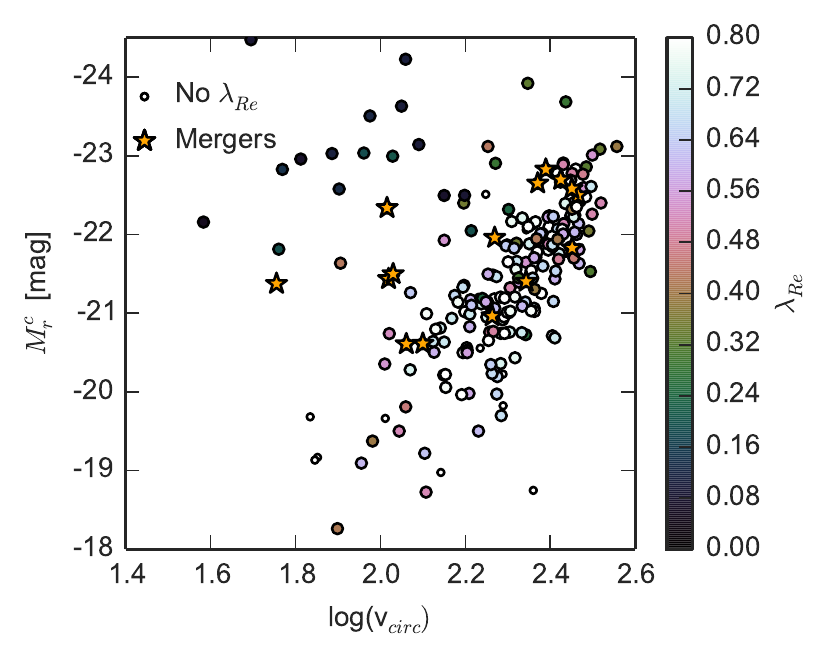}
\caption{Galaxies on the $v_{\mathrm{circ}} - M^c_r$ plane, colour-coded for inclination-corrected $\lambda_{\mathrm{Re}}$ parameter. The 20 galaxies without $\lambda_{\mathrm{Re}}$ values available (see text) are shown as smaller white circles. }
\label{fig:lambda_par}
\end{figure}

Fig. \ref{fig:lambda_par} shows that the galaxies on the circular velocity-luminosity plane are drawn from at least two parent distributions -- galaxies that exhibit significant ordered rotation and belong on the TFR, and the rest, including pressure-supported galaxies and some ongoing mergers. As a consequence, some of the data, even though they are of reasonable quality, are simply outside of the scope of the simple model of the TFR, which is a linear relation with small intrinsic scatter. Since linear regression is very sensitive to outliers and, more importantly, some of the galaxies in our sample do not belong on the Tully-Fisher relation by definition, some sort of outlier rejection must be performed.

\subsection{Modelling the TFR as a mixture of Gaussians}

\label{sec:MoG}
We did not apply any additional selection criteria to our galaxy sample (see Sec. \ref{sec:sample}), except for those that are implicit in the CALIFA mother sample selection and properties of the SDSS survey. As a result, it contains different galaxy populations, not all of which are well described by a thin rotating disk model assumed in Sec. \ref{sec:velfields} (mergers and slow rotators are two examples). 

We did not want to simply reject the outliers using an arbitrary procedure such as hand-pruning the data, sigma clipping or straightforward rejection of slow rotators and visually classified mergers. Instead, we modelled the distribution as a mixture of data obtained from two different generative models: a narrow, linear relation with Gaussian noise and small intrinsic scatter (corresponding to the subset of galaxies to which the TFR applies) and a two-dimensional Gaussian distribution that includes the galaxies that lie further away from the linear relation. 

The probability density function of a linear TFR with a small intrinsic Gaussian scatter $\sigma_i$ is

\begin{equation}
\begin{array}{l}
P(v|M, m, b, \sigma) = \dfrac{1}{\sqrt{2\pi( \sigma_{y}^2 + m^2 \sigma_{x}^2 + \sigma_{i}^2)}}\\\exp\left[-\dfrac{(M - mv - b)^2}{2( \sigma_{y}^2 + m^2 \sigma_{x}^2 + \sigma_{i}^2)}\right]
\end{array}
\end{equation}

\noindent where $v$ is the logarithm of circular velocity, M is the absolute magnitude, $m$ and $b$ are the slope and the offset of the linear relation.

The non-TF distribution is quite sparse, so we chose a non-restrictive two-dimensional Gaussian model described by its mean in two dimensions ($\mu_x, \mu_y$) and a covariance matrix 
\begin{equation}
\Sigma = \left| \begin{array}{cc}
\sigma_{\mathrm{bad_x}}^2 & \rho \sigma_{\mathrm{bad_x}}\sigma_{\mathrm{bad_y}}\\
\rho \sigma_{\mathrm{bad_x}}\sigma_{\mathrm{bad_y}} &  \sigma_{\mathrm{bad_y}}^2
\end{array} \right|
\end{equation}

Here $\sigma_{\mathrm{bad_{x, y}}}$ are the standard deviations of the non-TF points population, whose shape is allowed to vary, and $\rho$ is its correlation term.

If we combine both models, we can obtain a probability of belonging to the Tully-Fisher relation for each datapoint and reject the outliers based on this probability. We end up having 7 free parameters describing the two distributions ($m, b, \sigma_i, \mu_x, \mu_y, \sigma_{\mathrm{bad_x}}, \sigma_{\mathrm{bad_y}}$) which we infer and marginalise over the nuisance parameters $P_b$ (the probability of any point belonging to the non-TF distribution) by finding their posterior distributions using MCMC. The log-likelihood of the mixture of the two distributions described above is

\begin{equation}
\begin{array}{l}
  \ln\ \mathcal{L} \propto \\ \propto -0.5\Sigma\left((1 - P_b)
  \cdot\left[\ln(\sigma_{y}^2 + m^2 \sigma_{x}^2 + \sigma_{i}^2) +\dfrac{(y - mx -b)^2}{(\sigma_{y}^2 + m^2 \sigma_{x}^2 + \sigma_{i}^2)}\right]\right.\\\left. + P_b\cdot\left[\dfrac{(x - \mu_x)^2}{(\sigma_x^2 +\sigma_{\mathrm{bad_x}}^2)} + \dfrac{(y - \mu_y)^2}{(\sigma_y^2 +\sigma_{\mathrm{bad_y}}^2)}\right]\right)
\end{array}
\end{equation}

Modelling involves setting priors on several of the parameters. Due to the sparsity of the population offset from the TFR and the fact that we are working with the logarithm of velocity, which skews the error distribution, we apply Gaussian priors on its mean and variance, based on the estimated moments of the slow rotators population. We also apply a wide Gaussian prior, based on a simple linear fit to the fast rotators only, on the slope $m$, and, naturally, limit the $P_b$ to be between 0 and 1 and $\sigma_i, \sigma_{\mathrm{bad}} > 0$.

\begin{figure}[htbp]
\includegraphics[width=\linewidth]{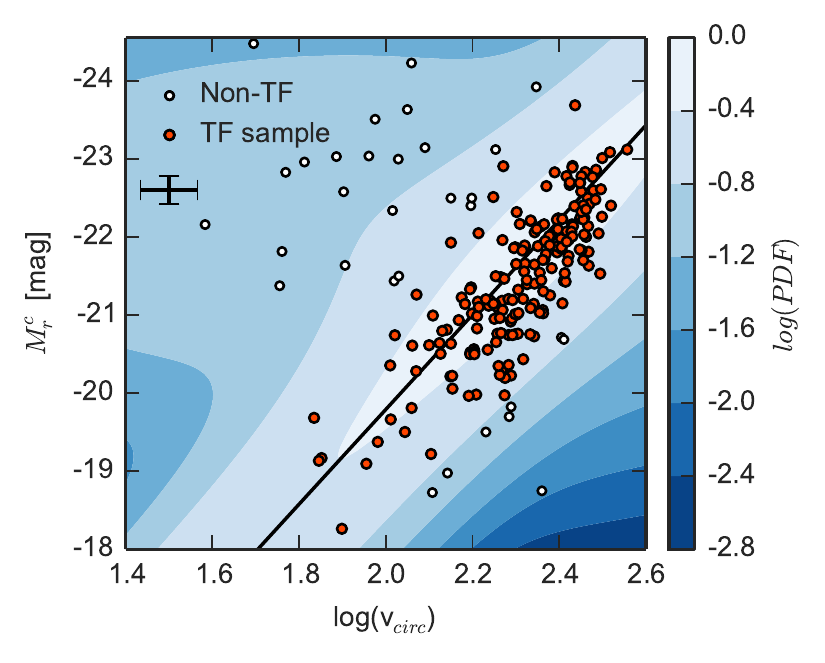}
\caption{Distribution of non-TF points (white) and the TF population (red), together with their underlying estimated generative functions (a 2D Gaussian and a linear relation with small intrinsic scatter), as described in the text. Error bars at the upper left corner show the mean uncertainties in $M^c_r$ and $\log(v_{\mathrm{circ}})$.}
\label{fig:MoGcontour}
\end{figure}

The results of the mixture modelling are shown in Fig. \ref{fig:MoGcontour}. We reject the datapoints with likelihoods lower than $1 - P_{\mathrm{good}} = 0.5$, i.e. the ones more likely to belong to the non-TFR distribution. This results in the rejection of 27 galaxies that are not compatible with being on a linear relation. The remaining 199 galaxies were used in further analysis and named the TF sample.

\subsection{Properties of the outlier galaxies}
\begin{figure}[!htbp]
\includegraphics[width=\linewidth]{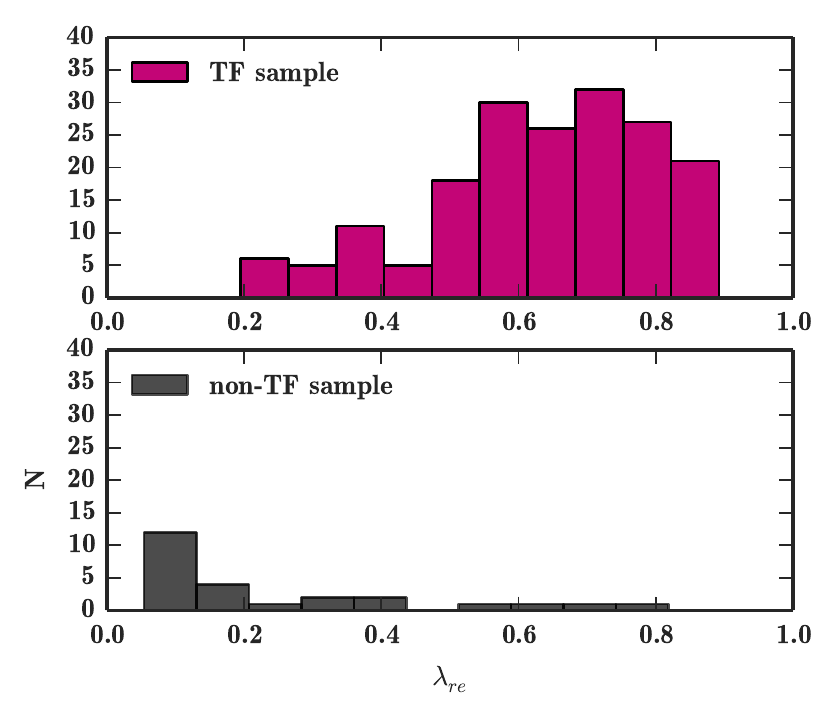}
\caption{ The lambda parameter $\lambda_{\mathrm{Re}}$ histogram for the TF sample (top) and the 25 outlier galaxies for which $\lambda_{\mathrm{Re}} $ values were available.}
\label{fig:lambda_hist}
\end{figure}

7 out of 27 rejected galaxies are slow rotators with $\lambda_{\mathrm{Re}}$ < 0.1 (Fig. \ref{fig:lambda_hist}). Two galaxies are classified as mergers, and several of the rejected galaxies are not sufficiently sampled by the CALIFA observations. 

\begin{figure}[htbp]
 \includegraphics[width=\columnwidth]{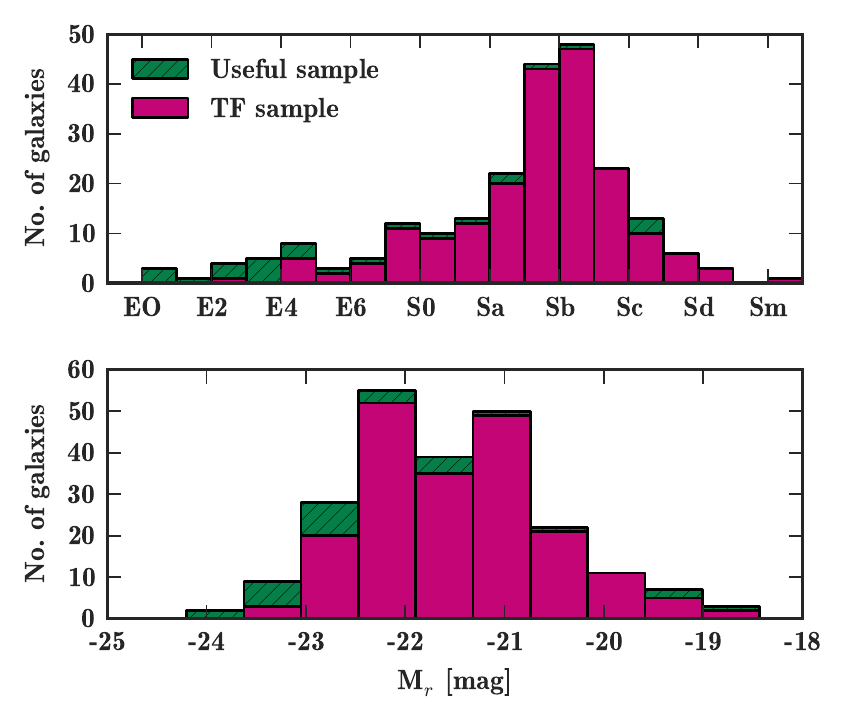}  
\caption{Morphological type and absolute magnitude histograms of the final TF sample and the useful sample.}
\label{fig:rejected_TF_samples}
\end{figure}

A comparison of absolute magnitudes and morphologies between the useful sample (described in Sec. \ref{sec:sample}) and the resulting Tully-Fisher and non-TF samples yielded by the mixture of Gaussians modelling is shown in Fig. \ref{fig:rejected_TF_samples}. The most salient property of the outlier rejection is the removal of the majority of bright early type ellipticals from the TF sample, which has a clear physical basis as such galaxies are much more likely to be slow rotators. However, even if this is expected, the rejection was based not on visual classification, but driven by the data, i.e. based on the location of the galaxy on the $M^c_r$ - $v_{\mathrm{circ}}$ plane and the uncertainties.

\subsection{Properties of the Tully-Fisher sample}

In order to characterise the final TF sample, we have compared the luminosity functions derived from the volume-corrected CALIFA mother sample and the volume-corrected TF sample (Fig. \ref{fig:LF_final}). We perform the procedure as described in Sec.\ref{sec:sample} and \citetalias{SCpaper}, by weighting each galaxy with its $1/V_{\mathrm{max}}$ factor. 

 They differ significantly at the lower luminosity end, where the LF of the TF sample falls off sharply. The difference is not so pronounced for the brightest galaxies, due to the low number statistics for such objects in the volume-complete sample. Note that we did not \textit{a priori} expect to retain the volume completeness during the outlier rejection, as the rejection is non-random. The difference between the luminosity functions is an expected outcome of the outlier rejection procedure described in Sec. \ref{sec:MoG}.

However, the mixture modelling is a reproducible procedure. Given a statistically representative sample, the same procedure can be performed again, yielding a distribution of galaxies that is representative of the overall rotation-supported galaxy population and is a subset of the joint volume-complete sample (within the limits of the observed sample). Even so, Fig. \ref{fig:LF_final} shows that the Tully-Fisher sample can nevertheless be considered to be volume complete-able within the -20 $\geq M^c_r \geq$ -22 magnitude range.

\begin{figure}[!htbp]
\includegraphics[width=\linewidth]{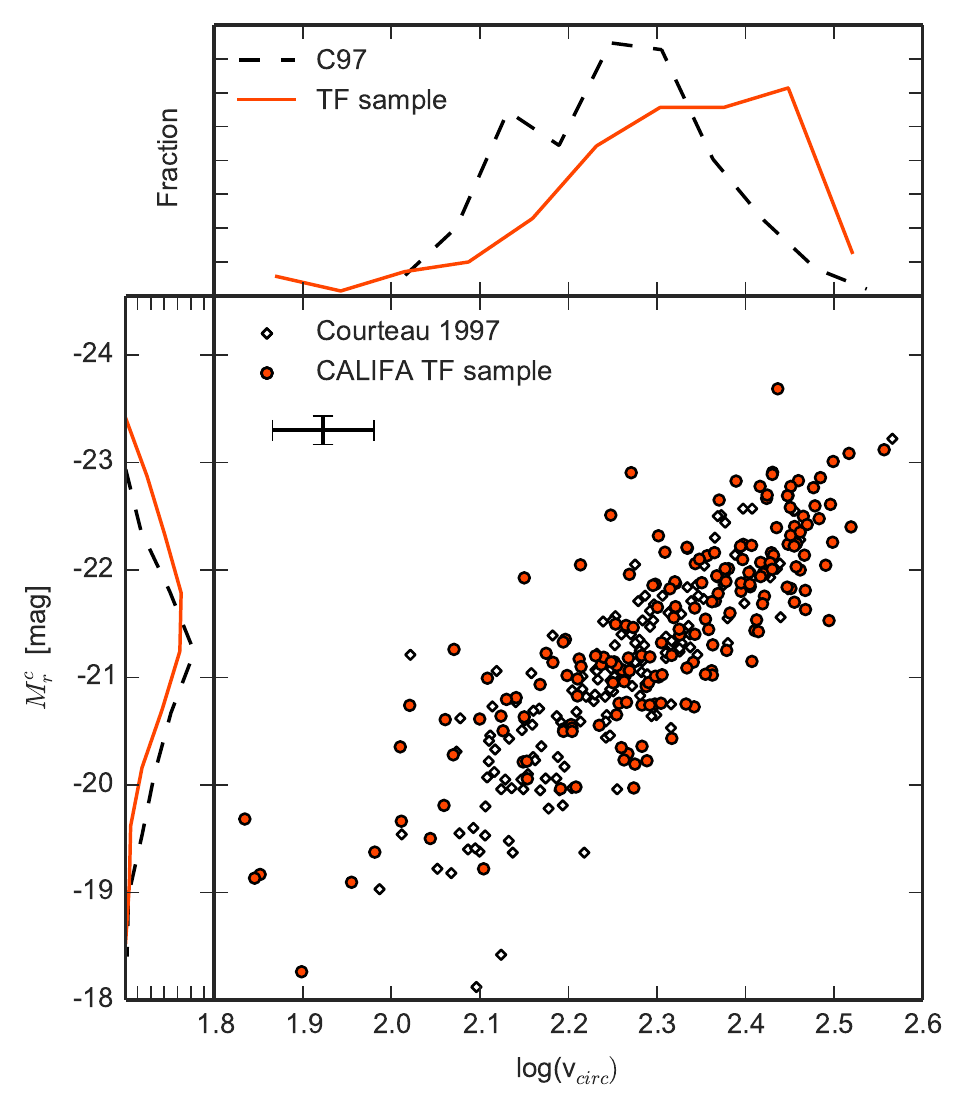} 
\caption{Comparison of our Tully-Fisher sample $M^c_r$ - $v_{\mathrm{circ}}$ distribution and \citet{Courteau1997} (C97) $M_r^c$ - H$\alpha$ $v_{\mathrm{opt}}$ measurements. The marginal plots show normalised histograms for both samples. The mean uncertainty magnitudes of CALIFA measurements are shown by the error bars.}
\label{fig:TF_Courteau_marginal}
\end{figure}

In order to check that velocity field-based measurements do not show systematic offsets from the conventional long-slit measurements, we show the $M^c_r$ - $v_{\mathrm{circ}}$ distribution of the TF sample and a comparison sample from \citet[hereafter C97]{Courteau1997} in Fig. \ref{fig:TF_Courteau_marginal}. The TF sample shows a larger scatter than the comparison sample (the $rms$ error of log($v_{\mathrm{circ}}$) is equal to 0.26 dex), which is expected given that the comparison sample used a sample of galaxies with specific selection criteria, such as late Hubble types, moderately high inclination and lack of interactions or peculiar properties \citep{Courteau1996}. The TF sample has a higher proportion of brighter galaxies of earlier types.
In addition to that, the CALIFA stellar circular velocities are typically larger than H$\alpha$ $v_{\mathrm{opt}}$ measurements due to the circular velocity correction. A direct comparison would involve calculating an equivalent correction for the gas, which is not a trivial endeavour as discussed in Sec.\ref{ADC}. Nevertheless, the plot shows that circular velocity measurements from IFS velocity fields and rotation curves are compatible and do not show a significant systematic deviation from a similar underlying relation.

\section{The Tully-Fisher relation}
\label{sec:TF}
Even though a simple straight-line model is not accurate, it is a useful tool in the area of distance determination and has been widely used to model the TFR, as well as to compare the local relation with high-redshift galaxy samples. 

For comparison purposes we fit the TFR as a straight line with free slope, intrinsic scatter and offset (zeropoint) parameters $m, \sigma_i, b$, assuming $M^c_r$ as the independent variable. We use the \textsc{HYPER-FIT} hyperplane fitting package \citep{Robotham2015} which provides the tools to fit heteroscedastic and covariant data, and use $1/V_{\mathrm{max}}$ values as the fit weights. 

\begin{figure}[!htbp]
\includegraphics[width=\linewidth]{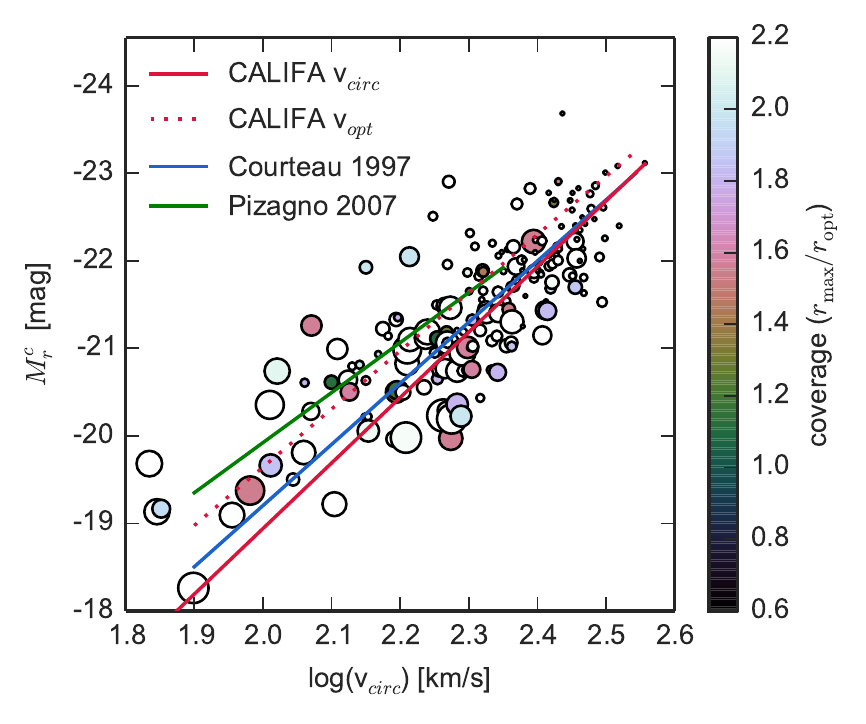} 	
\caption{Linear fit (see text for more details) to CALIFA $v_{\mathrm{circ}}$ - $M^c_r$ data, and comparison with \citet{Courteau1997} and \citet{Pizagno2007} TF fits. Point sizes are proportional to $1/V_{\mathrm{max}}$ weights, colours indicate insufficient spatial coverage of the velocity field (sampling of the rotation curve is lower than 1 $r_{\mathrm{opt}}$). The fit to $v_{\mathrm{opt}}$ -  $M^c_r$ is also shown on the plot.}
\label{fig:TFR}
\end{figure}

In order to check if the limited spatial coverage of CALIFA velocity fields and the necessity to extrapolate some rotation curves beyond the last measured point (see Sec.\ref{sec:sample}) did not introduce a bias, we check where our datapoints lie on the Tully-Fisher relation and do not notice a significant offset (Fig. \ref{fig:TFR}).

We have also made sure that the infall corrections described in Sec. \ref{sec:lumdata} did not affect the absolute magnitudes significantly by comparing them with luminosities estimated using pure Hubble-flow based distances. We find that although the average difference between the two values is equal to 0.14 mag, there is no systematic effect on the TFR. Similarly, the lack of the extinction corrections described in Sec. \ref{sec:lumdata} had a negligible effect on the intrinsic scatter of the TFR, however, it made the slope slightly steeper.

We compare our fit result with \citetalias{Courteau1997} and \citet[hereafter P07]{Pizagno2007} who investigate the $r$-band TF relation using H$\alpha$ rotation curves, based on $v_{\mathrm{opt}}$. Our TFR, fit using a $v_{\mathrm{circ}}$ estimate, is shifted to the right and is steeper. Using the $1/V_{\mathrm{max}}$ weights leads to a slightly flatter TFR. This is not surprising, given that the unweighted relation is dominated by the more luminous galaxies. Applying volume weights acts in the opposite direction and brings the fit relation closer to \citetalias{Courteau1997} results which are based on a sample dominated by late-type spirals as shown in Fig.\ref{fig:TF_Courteau_marginal}. 

The slope, offset and scatter values for $v_{\mathrm{circ}}$ and $v_{\mathrm{opt}}$-based Tully-Fisher relation respectively are provided in Table \ref{table:TF_params}. The table also contains the forward TFR parameters provided by \citetalias{Courteau1997} and \citetalias{Pizagno2007}. We note that the scatter value reported by us is not the standard deviation of the points from the straight fit line, but the intrinsic scatter not accounted for by the measurement uncertainties during the modelling. The $rms$ error of our measurements is 0.26 mag.

The intrinsic scatter value we obtain suggest low upper limits on the intrinsic scatter of the TFR. The sources of it include scatter in the dark matter halo spin, concentration and response to galaxy formation \citep{Dutton2011}, potential ellipticity \citep{Franx1992} and formation history \citep{Eisenstein1996, Giovanelli1997}, mass-to-light ratio \citep{Gnedin2007} and morphology \citep{Giovanelli1997}, among others. A more in-depth study of the intrinsic TFR scatter that would include the additional measurement errors resulting from our adopted infall, extinction and circular velocity corrections is outside of the scope of this paper, but it is unlikely that the reported intrinsic scatter would be increased.
\begin{table}
\caption{Tully-Fisher relation fit parameters and literature values. }              
\label{table:1}      
\centering                                      
\begin{tabular}{c c c c}          
\hline\hline                        
Velocity definition & slope & offset & scatter \\    
\hline                                   
   CALIFA $v_{\mathrm{circ}}$ & -7.5$\pm$0.5 & -4.0$\pm$1.0 & 0.03$\pm$0.06\\      
   CALIFA $v_{\mathrm{opt}}$ & -6.7$\pm$0.4 & -6.3$\pm$0.9 & 0.09$\pm$0.03\\
  \citetalias{Courteau1997} $v_{\mathrm{opt}}$ & -6.99$\pm$0.33 & -5.23$\pm$0.46 & 0.46\\
   \citetalias{Pizagno2007} $v_{\mathrm{opt}}$ & -5.72$\pm$0.19 & -7.9$\pm$0.03 & 0.42\\

\hline                                             
\end{tabular}
\label{table:TF_params}
\end{table}

\section{Volume-corrected bivariate distribution function in the Tully-Fisher plane}
\label{sec:biv_distr}

It is possible to use the volume and large-scale correction procedure described in Sec. \ref{sec:sample} to reconstruct a volume-complete bivariate distribution in the $M^c_r-v_{\mathrm{circ}}$ plane, applicable within the CALIFA completeness limits. 

We use kernel density estimation (KDE) to achieve this. KDE is a non-parametric probability density estimation procedure \citep{rosenblatt1956, parzen1962} consisting of representing each datapoint as a smooth distribution and then inferring the underlying density distribution. It is superior to histograms because a smooth kernel can be chosen, there is no dependence on the choice of the starting bin and the probability density function is obtained naturally.

\begin{figure*}[ht]
\includegraphics[width=\columnwidth]{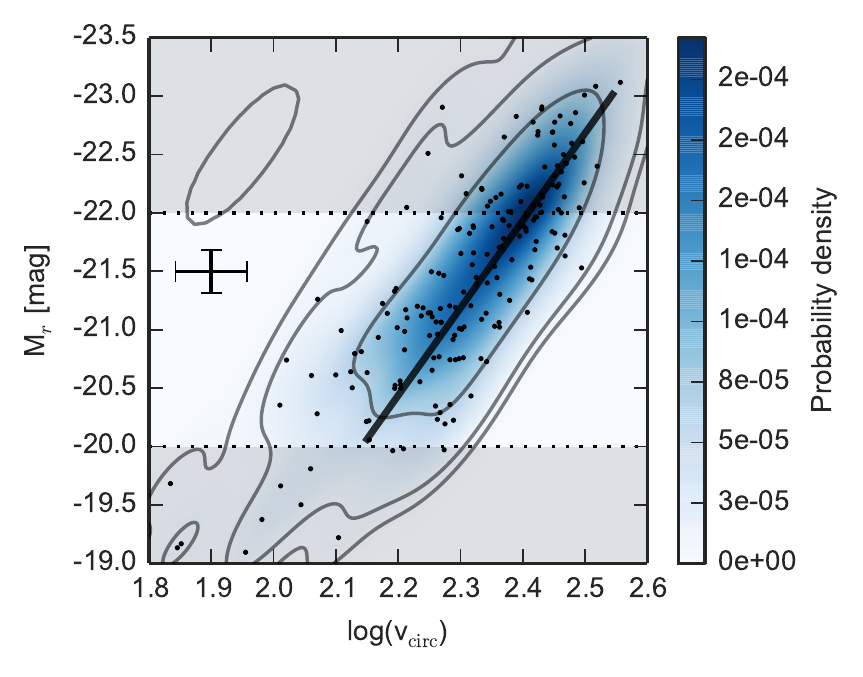} 
\includegraphics[width=\columnwidth]{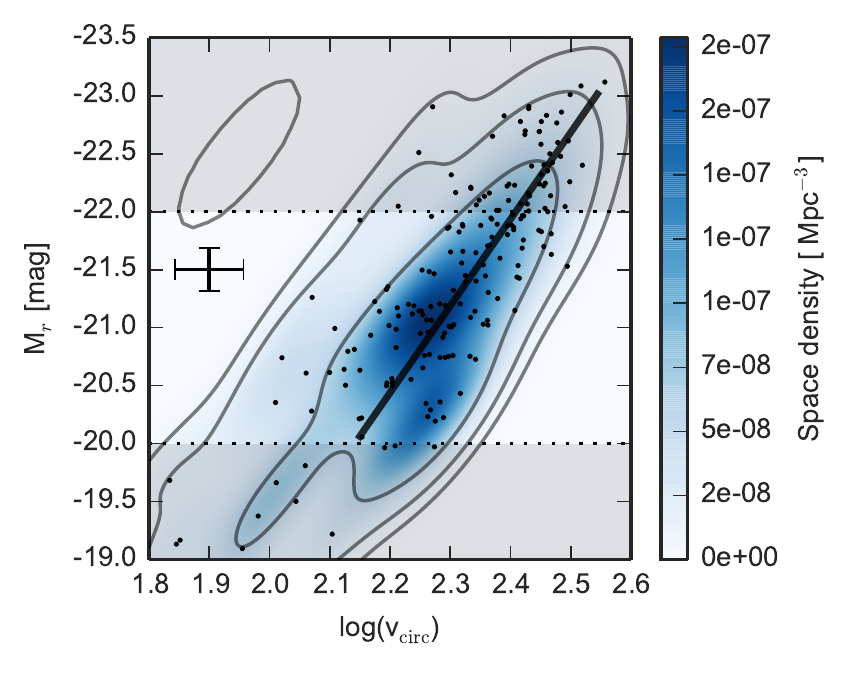} 
\caption{The left plot shows the joint probability density of $M^c_r$-$v_{\mathrm{circ}}$. The right plot shows the joint space densities distribution, estimated by weighting the KDE kernels by the $1/V_{\mathrm{max}}$ weights. Grey lines denote the 1, 2, and 3 standard deviation contours. The black line shows the Tully-Fisher relation discussed in the previous section. A 1-$\sigma$ contour of the Gaussian kernel used for density estimation is shown at the upper left corners. Grey shaded regions mark the regions where our sample is not complete.} 
\label{fig:TF_KDE}
\end{figure*}

We chose a two-dimensional Gaussian as the kernel function. Selecting the size and orientation (the kernel function's covariance matrix) can be done using various methods, such as cross-validation, or using one of the rules-of-thumb that empirically estimate the bandwidth based on the number of data points and dimensions of the dataset. 
We estimate the optimal kernel bandwidth based on the global shape of the distribution and the Silverman's rule \citep{Silverman1986}. First of all, the global covariance matrix of all the observed points is estimated, assuming that the observed points distribution is similar to a Gaussian distribution in a sense that it is unimodal, symmetric and not heavy-tailed. Then this matrix is multiplied by a scaling factor $f_s = n^{-\frac{1}{6}}$, derived according to the Silverman's rule \citep{Scott2005}. Here $n$ is the number of data points. 

We use the \textit{stats.gaussian_kde} routine from \textit{SciPy} package \citep{SciPy} as the basis for our analysis. The left panel of Fig. \ref{fig:TF_KDE} shows the probability density distribution in the $v_{\mathrm{circ}}$ - $M^c_r$ space. The density distribution here is dominated by the brighter galaxies with $M^c_r$ > -20.5 mag and $v_{\mathrm{circ}} $ > 200 km/s due to the CALIFA sample construction and smaller relative scatter at larger $\log(v_{\mathrm{circ}})$ values.

However, this picture changes when the $1/V_{\mathrm{max}}$ factors are included as additional KDE weights, as shown in the right panel of Fig. \ref{fig:TF_KDE}. The area with the highest probability density now shifts to lower velocities (< 250 km/s) and magnitudes ($M^c_r$ > -21.5 mag). We convert the probability density to space densities by multiplying the probability density (which integrates to 1) by the sum of all the $1/V_{\mathrm{max}}$ factors in the TF sample. 

The joint distribution of the luminosity function and the velocity function (discussed in an upcoming paper) could constrain galaxy formation and evolution models more than a single marginal distribution (LF or VF). The linear Tully-Fisher relation does not directly provide information about the number of galaxies at a given location in the $M^c_r-v_{\mathrm{circ}}$ plane, whereas we provide space densities which could be compared with simulations of cosmological volumes.

When comparing a model or simulation of a galaxy with the Tully-Fisher relation, traditionally it consists of making sure that the produced galaxies lies on the TFR. The TFR is typically defined by the slope and offset parameters, in some cases including the scatter. Here we have determined the full volume-corrected bivariate distribution, or probability distribution, in the $L-v_{\mathrm{circ}}$ plane for the first time. This allows a much more direct and quantitative estimate of the likelihood that a simulated galaxy is consistent with the real galaxy population. The halo velocities obtained from cosmological simulations would have be converted into the circular velocity, which can be directly compared with our results. Analysis of such a volume-complete distribution, however requiring a larger sample spanning diverse environments and luminosities, could shed some light on environmental influence on the TFR, indicated by \citet{Blanton2008}.

Our analysis is limited by incompleteness issues at the low velocity/fainter magnitude end. Nevertheless, the difference in the two distributions is apparent, showing that the most luminous galaxies do not contribute significantly to the bulk of the Universe's stellar angular momentum. The space densities shown in Fig. \ref{fig:TF_KDE}, as well as the other data, are available online at CDS.

\section{Conclusions}
In this paper we present the first space density distribution of rotating galaxies in the $M^c_r$-$v_{\mathrm{circ}}$ plane, derived using the CALIFA stellar velocity fields. The use of stellar IFS kinematics, careful extinction corrections and the statistically well-understood CALIFA sample allows us to perform volume corrections and provide a fair representation of the distribution of galaxies with -20  $> M^c_r >$ -22 mag. Our key messages and results are as follows.
\begin{itemize}
  \item We find that the velocity uncertainties in many TFR analyses are underestimated. The reason for this is a combination of direct use of photometric inclination estimates, lack of full 2D spatial information and degeneracies between rotation curve parameters and inclination. Using consistent MCMC modelling of velocity fields we obtain realistic velocity uncertainties, which are propagated to the further analysis.
  \item By avoiding any arbitrary cuts in our sample and instead modelling the TFR and non-TFR populations of galaxies we are using a reproducible, probabilistic approach to outlier rejection (Sec.\ref{sec:MoG}). This allows us to generalise our analysis to other samples and lets us preserve the capability to perform volume corrections.
    
     \item A $1/V_{\mathrm{max}}$-weighted linear fit with bivariate uncertainties provided an $r$-band TFR with slope, zeropoint and scatter equal to -7.5, -4.0, 0.03 for the $v_{\mathrm{circ}}$-based TFR and -6.7, -6.3 and 0.09 for $v_{\mathrm{opt}}$-based Tully-Fisher relation (Sec. \ref{sec:TF}).
    
  \item We provide a bivariate local space density distribution in the $v_{\mathrm{circ}} - M^c_r$ plane (Sec. \ref{sec:biv_distr}), which, although less straightforward to compare with than a simple linear parameterisation, provides more information than a single line and is more representative of the overall properties of galaxies. The full two-dimensional distribution is what simulations could be compared with. 
\end{itemize}

\begin{acknowledgements}

We sincerely thank S. Courteau and the anonymous referee for carefully reading the draft, thoughtful suggestions and providing us with an opportunity to improve the manuscript.

SB acknowledges financial support from BMBF through the Erasmus-F project (grant number 05 A12BA1). CJW acknowledges support through the Marie Curie Career Integration Grant 303912. JFB. acknowledges support from grant AYA2013-48226-C3-1-P from the Spanish Ministry of Economy and Competitiveness (MINECO). IM acknowledges financial support from the Spanish Ministry of Economy and Competetiveness through the grant A\&A2013-0422277.

This study makes uses of the data provided by the Calar Alto Legacy Integral Field Area (CALIFA) survey (\url{http://califa.caha.es/}). It is based on observations collected at the Centro Astronómico Hispano Alem\'{a}n (CAHA) at Calar Alto, operated jointly by the Max-Planck-Institut fűr Astronomie and the Instituto de Astrofisica de Andalucia (CSIC). 

In addition to that, we used the data from SDSS DR7 \citep{SDSSDR7}. Funding for the SDSS and SDSS-II has been provided by the Alfred P. Sloan Foundation, the Participating Institutions, the National Science Foundation, the U.S. Department of Energy, the National Aeronautics and Space Administration, the Japanese Monbukagakusho, the Max Planck Society, and the Higher Education Funding Council for England. The SDSS Web Site is \url{http://www.sdss.org/}.

The SDSS is managed by the Astrophysical Research Consortium for the Participating Institutions. The Participating Institutions are the American Museum of Natural History, Astrophysical Institute Potsdam, University of Basel, University of Cambridge, Case Western Reserve University, University of Chicago, Drexel University, Fermilab, the Institute for Advanced Study, the Japan Participation Group, Johns Hopkins University, the Joint Institute for Nuclear Astrophysics, the Kavli Institute for Particle Astrophysics and Cosmology, the Korean Scientist Group, the Chinese Academy of Sciences (LAMOST), Los Alamos National Laboratory, the Max-Planck-Institute for Astronomy (MPIA), the Max-Planck-Institute for Astrophysics (MPA), New Mexico State University, Ohio State University, University of Pittsburgh, University of Portsmouth, Princeton University, the United States Naval Observatory, and the University of Washington.

This research has made use of the NASA/IPAC Extragalactic Database (NED), which is operated by the Jet Propulsion Laboratory, California Institute of Technology, under contract with the National Aeronautics and Space Administration, and of NASA's Astrophysics Data System Bibliographic Services.

We have extensively used and are grateful for the open source data analysis and visualisation tools: \textit{Matplotlib} \citep{Matplotlib}, \textit{SciPy} \citep{SciPy}, \textit{hyper-fit} \citep{Robotham2015} and \textit{prettyplotlib} (\url{https://github.com/olgabot/prettyplotlib}). 
\end{acknowledgements}

\bibliographystyle{aa}
\bibliography{tf}

\end{document}